\documentclass[journal]{IEEEtran}
\usepackage{cite}
\usepackage{amsmath,amssymb,amsfonts}
\usepackage{textgreek}
\usepackage{algorithmic}
\usepackage{graphicx}
\usepackage[caption=false,font=normalsize,labelfont=sf,textfont=sf]{subfig}
\usepackage{textcomp}
\usepackage{flushend}
\usepackage{xcolor}
\usepackage{relsize}
\def\BibTeX{{\rm B\kern-.05em{\sc i\kern-.025em b}\kern-.08em
    T\kern-.1667em\lower.7ex\hbox{E}\kern-.125emX}}

\usepackage{tipa}
\usepackage[linesnumbered,ruled,vlined]{algorithm2e}

\begin{document}
\pagenumbering{gobble}
\title{\huge{Data-Aided Variational Bayesian Inference for CSI Estimation over Doubly-Selective DCO-OTFS MIMO VLC Systems with Affine-Precoded Superimposed Training Sequences}}

\author{
Shubham~Saxena,~\IEEEmembership{Graduate Student Member,~IEEE,} Suraj~Srivastava,~\IEEEmembership{Member,~IEEE,} and Aditya~K.~Jagannatham,~\IEEEmembership{Senior Member,~IEEE}\vspace{-5mm} 
\thanks{Shubham Saxena, and Aditya K. Jagannatham are with the Department of Electrical Engineering, Indian Institute of Technology Kanpur, Kanpur-$208016$, India (e-mail: \{shubs20;  adityaj\}@iitk.ac.in). Suraj Srivastava is with the Department of Electrical Engineering, Indian Institute of Technology Jodhpur, Rajasthan $342030$, India (email: surajsri@iitj.ac.in).
}
}

\maketitle

\begin{abstract}
An orthogonal affine-precoded superimposed training sequences (AP-STS)-based architecture is developed for the cyclic prefix
(CP)-aided multiple input multiple output (MIMO) direct-current-biased orthogonal time frequency space
(DCO-OTFS) visible light communication (VLC) systems relying on arbitrary transmitter-receiver pulse shaping. The data and pilot symbol matrices are
affine-precoded (AP) and superimposed in the delay-Doppler (DD)-domain for each transmit light-emitting diode (LED), followed by the development of an end-to-end DD-domain relationship for the input-output symbols. At the receiver for each receiver photodiode (PD), the decoupled pilot and data symbol are extracted by employing orthogonal precoder matrices, which eliminates the mutual interference. Furthermore, a novel pilot-aided (PA) variational Bayesian inference (PA-VBI) technique is conceived for the channel state information (CSI) estimation of MIMO DCO-OTFS VLC systems based on the expectation-maximization (EM) technique. Subsequently, a data-aided (DA) variational Bayesian inference (DA-VBI)-based joint CSI estimation and data detection technique is proposed, which beneficially harnesses the estimated data symbols for improved CSI estimation. 
Moreover, the Bayesian Cramer-Rao lower bounds (BCRLBs) are also derived for MIMO DCO-OTFS VLC systems. Finally, simulation results demonstrate that the proposed method yields superior performance in terms of normalized mean-square-error (NMSE), pilot overhead, and symbol
error-rate (SER).
\end{abstract}

\vspace{-3mm}

\section{Introduction}

\IEEEPARstart{E}{scalating} requirements for high-throughput mobile services, fueled by pervasive Internet of Things (IoT) deployments and smartphone-driven applications, have further strained the limited radio-frequency (RF) spectrum, exacerbating spectrum scarcity. As a complementary access paradigm, visible light communication (VLC) has therefore received growing attention \cite{h1}. VLC benefits from the availability of a broad, unlicensed optical band and from the maturity of low-cost optoelectronic hardware such as light-emitting diodes (LEDs) and photodiodes (PDs), which enables simultaneous illumination and data transmission. In addition, VLC exhibits low electromagnetic radiation and strong immunity to electromagnetic interference. Since optical signals are largely confined to indoor environments and do not penetrate opaque obstacles, VLC can also provide inherent resistance to external eavesdropping. Moreover, reusing existing lighting infrastructure can lower deployment cost and energy consumption, supporting energy-efficient communication \cite{h1,8123892,xu2023optical,saxena2023sparse}.

VLC links typically exhibit a composite propagation structure comprising a dominant line-of-sight (LoS) path together with multiple non-LoS (NLoS) components. The LoS term represents the direct optical coupling between the LED transmitter and the PD, whereas the NLoS terms are generated by reflections from walls, furniture, and other indoor surfaces. The characteristics of multipath VLC channels have been investigated in \cite{chen2016adaptive,schulze2016frequency}. Moreover, the influence of higher-order reflections was examined in \cite{zhou2014impact}, demonstrating that models restricted to only a small number of reflection paths may not adequately capture the channel behavior for high-throughput VLC operation. Along similar lines, \cite{saxena2025multiple} adopts a multipath VLC description that explicitly incorporates both LoS and NLoS contributions. 
The resulting delay dispersion can lead to inter-symbol interference (ISI) in indoor VLC scenarios \cite{chen2016adaptive,schulze2016frequency,zhou2014impact,saxena2025multiple}. To mitigate ISI, optical orthogonal frequency division multiplexing (O-OFDM) has been extensively explored as a multicarrier solution for LED-based VLC, providing high spectral efficiency while offering resilience against multipath-induced distortion \cite{van2021deep,h1}. In addition, owing to its low-complexity hardware requirements, intensity modulation with direct detection (IM/DD) is widely employed for O-OFDM reception \cite{h1,g1}.

In parallel, orthogonal time frequency space (OTFS) modulation has emerged as a promising technique for communication over doubly selective channels encountered in mobile environments \cite{hadani2017orthogonal}. In contrast to conventional orthogonal frequency division multiplexing (OFDM), which is vulnerable to Doppler-driven intercarrier interference (ICI), OTFS maps information symbols onto a delay-Doppler (DD) lattice, thereby enhancing robustness to time-frequency (TF) selectivity and facilitating more dependable channel estimation (CE). While DD-domain signal processing generally increases implementation complexity, OTFS has been reported to retain notable performance gains even in quasi-static multipath settings \cite{raviteja2019otfs}. These findings have stimulated interest in extending OTFS to the optical domain, leading to optical OTFS (O-OTFS) frameworks for VLC \cite{zhong2020orthogonal,zheng2021dco}. In such systems, the achievable throughput is highly sensitive to the fidelity of channel state information (CSI). Since classical CE approaches may impose substantial pilot overhead and may be less reliable in doubly selective optical channels, recent studies have emphasized more efficient CSI acquisition mechanisms. The most relevant works are discussed next.

\vspace{-4mm}
\subsection{Literature Review}
Since VLC relies on IM/DD, the transmitted signal must be real and non-negative, thereby requiring appropriate optical adaptations of OTFS. Zhong \textit{et al.} \cite{zhong2020orthogonal} introduced direct current-biased O-OTFS (DCO-OTFS) by imposing two-dimensional Hermitian symmetry, and reported improved bit error-rate (BER) and reduced peak-to-average power ratio (PAPR) relative to direct current-biased O-OFDM (DCO-OFDM). Zheng \textit{et al.} \cite{zheng2021dco} integrated DCO-OTFS within a full-duplex, relay-assisted VLC setup and showed that cyclic-prefix (CP) overhead can be lowered compared with DCO-OFDM, yielding higher spectral efficiency. To account for optical front-end impairments, Sharma \textit{et al.} \cite{sharma2023hyperparameter} proposed a hyperparameter-free receiver using random Fourier features to mitigate LED nonlinearity in O-OTFS VLC links. Sinha \textit{et al.} investigated O-OTFS for static indoor dual-LED VLC in \cite{sinha2021otfs} and extended the study to quad-LED configurations in \cite{sinha2021quad}; both works demonstrated performance gains over O-OFDM. Xu \textit{et al.} \cite{xu2023optical} developed an O-OTFS framework highlighting bandwidth, power, and energy-efficiency advantages while enabling DD-domain CSI estimation. In a related direction, Wang \textit{et al.} \cite{wang2025spectrally} derived a general DD to time-domain (TD) conversion via the discrete Zak transform (DZT) and formulated a DD-domain maximum likelihood detector. More recently, Cai \textit{et al.} \cite{cai2025power} proposed a power- and spectrum-efficient O-OTFS scheme for underwater VLC, whereas Chen \textit{et al.} \cite{chen2025optical} examined an IM/DD satellite optical link employing DCO-OTFS to counteract fading and Doppler effects.


\begin{table*}[t]
\centering
\caption{\textcolor{black}{Boldly contrasting our contributions to the literature}}
\label{SoTA}
\resizebox{\linewidth}{!}{%
\begin{tabular}{|l|c|c|c|c|c|c|c|c|c|c|c|c|c|c|c|}
\hline
\textbf{Features} & \cite{xu2023optical} & \cite{saxena2023sparse} & \cite{zhong2020orthogonal} & \cite{zheng2021dco} & \cite{sharma2023hyperparameter} & \cite{sinha2021quad} & \cite{wang2025spectrally} & \cite{cai2025power} & \cite{chen2025optical} & \cite{liao2023sparse} & \cite{mishra2021iterative} & \cite{yuan2021data} & \cite{estrada2019superimposed} & \cite{wang2022joint} & \textbf{Proposed} \\
\hline
O-OTFS & \checkmark &  & \checkmark & \checkmark & \checkmark & \checkmark & \checkmark & \checkmark & \checkmark & \checkmark &  &  &  &  & \checkmark \\ \hline
DD-domain sparsity &  &  &  &  &  &  &  &  &  & \checkmark & \checkmark & \checkmark &  & \checkmark & \checkmark \\ \hline
STS &  &  &  &  &  &  &  &  &  &  & \checkmark & \checkmark & \checkmark &  & \checkmark \\ \hline
Data-aided CE &  &  &  &  &  &  &  &  &  &  & \checkmark & \checkmark &  & \checkmark & \checkmark \\ \hline
MIMO VLC &  &  &  &  &  & \checkmark &  &  &  &  &  &  & \checkmark &  & \checkmark \\ \hline
VBI-based CE &  &  &  &  &  &  &  &  &  &  &  &  &  & \checkmark & \checkmark \\ \hline
BCRLB &  & \checkmark &  &  &  &  &  &  &  &  &  &  &  &  & \checkmark \\ \hline
Affine-Precoding &  &  &  &  &  &  &  &  &  &  &  &  &  &  & \checkmark \\ \hline
Fractional Doppler &  &  &  &  &  &  &  &  &  &  &  &  &  &  & \checkmark \\ \hline
\end{tabular}%
}
\vspace{-2mm}
\end{table*}

Accurate CSI acquisition is crucial in VLC since the coexistence of LoS and NLoS paths typically results in delay dispersion, while user or platform motion may introduce Doppler spread, together creating doubly selective channels that complicate both estimation and detection. For O-OFDM, Saxena \textit{et al.} \cite{saxena2023sparse} proposed Bayesian learning (BL)-based CSI estimation for frequency-selective DCO-OFDM and asymmetrically clipped O-OFDM (ACO-OFDM) VLC links, reporting notable improvements over conventional estimators. In O-OTFS, Liao \textit{et al.} \cite{liao2023sparse} derived a DD-domain pilot-to-channel input-output relation and developed an embedded-pilot BL estimator. Nevertheless, embedded designs still consume resources for pilots and protection regions, motivating more bandwidth-efficient training, particularly for multiple-input multiple-output (MIMO) VLC. Complementary RF-OTFS studies highlight that impulse-type pilot schemes can yield interference-free DD observations via antenna-wise pilot orthogonalization, but often require large guard regions and threshold tuning, increasing overhead and reducing robustness \cite{ramachandran2018mimo}.
Embedded pilot-data structures offer a more compact alternative by sharing a single OTFS frame, yet the accompanying pilot placement and guard requirements can still cause noticeable throughput loss \cite{hadani2017orthogonal}. A salient property of OTFS channels is their inherent sparsity in the DD-domain due to the limited number of dominant scatterers, which can be leveraged to improve CSI acquisition accuracy \cite{li2020new,li2023doubly,srivastava2022otfs}. In this context, Li et al. \cite{li2020new} developed a modified orthogonal matching pursuit (OMP) strategy to exploit DD sparsity, while Li and Yu \cite{li2023doubly} proposed a graph-based turbo  minimum mean-square-error (MMSE) method that does not explicitly assume sparsity. Although such approaches substantially outperform conventional MMSE baselines, BL methods frequently achieve superior performance by more effectively capturing sparse structure \cite{saxena2023sparse,srivastava2022otfs}. For example, \cite{srivastava2022otfs} presented a BL-based estimator applicable under arbitrary transmit and receive pulse shaping. Nevertheless, BL techniques are iterative and can be computationally demanding because each iteration involves inversion of high-dimensional matrices that couple sparse channel coefficients with the sensing dictionary \cite{saxena2023sparse,srivastava2022otfs}. Beyond sparsity-centric schemes, Liu et al. \cite{liu2020uplink} investigated uplink-aided downlink estimation and employed an expectation maximization variational Bayesian inference (VBI) procedure to infer uplink channel parameters. Notably, these methods primarily rely on pilots, whereas our premise is that additional gains are achievable by also incorporating data symbols in the estimation process.

To enhance training efficiency, superimposed training sequence (STS) methods overlay pilot symbols directly on top of data symbols \cite{mishra2021iterative,yuan2021data,muntane2024optimal}. In VLC, Estrada \textit{et al.} \cite{estrada2019superimposed} examined least squares (LS)-based CSI estimation for a multiple-input single-output (MISO) DCO-OFDM configuration employing straightforward pilot-data superposition. However, such direct overlay unavoidably introduces mutual interference between pilots and data. Suppressing this impairment often relies on maximum \textit{a posteriori} processing or message-passing based detectors and typically presumes prior knowledge (e.g., channel order or sparsity level), which may be difficult to justify in practice \cite{mishra2021iterative,yuan2021data}. Affine-precoded STS (AP-STS) offer a more structured approach by using orthogonal affine precoders, enabling algebraic projection at the receiver to separate pilot and data components and, ideally, to avoid pilot-data interference across a wide SNR range. Although affine precoding has been explored in RF scenarios \cite{tran2008orthogonal}, most existing developments primarily target time-invariant channels and do not explicitly accommodate doubly selective behavior, nor the IM/DD requirements and optical front-end characteristics that are fundamental to VLC. Moreover, AP-STS can be further reinforced through data-aided processing, wherein detected symbols are reused as reliability-weighted virtual pilots to iteratively refine CSI within a BL-based learning framework.

Building on the above limitations, this work proposes a data-aided VBI framework for CSI estimation in doubly selective DCO-OTFS MIMO VLC systems with AP-STS signaling. The proposed approach exploits DD-domain sparsity stemming from a small set of dominant reflectors, thereby enhancing estimation fidelity without assuming prior statistical knowledge of the channel order or sparsity level. VBI further enables tractable approximation of otherwise intractable posterior distributions by adopting simplified surrogate distributions, which promotes faster convergence and reduced computational burden. By jointly utilizing interference-free pilot separation enabled by orthogonal affine precoding and probabilistically weighted data-aided refinements, the proposed scheme attains improved normalized mean-square-error (NMSE) and symbol error-rate (SER) while also reducing pilot overhead. Moreover, the framework is benchmarked against standard sparse recovery baselines, including OMP and FOCal Underdetermined System Solver (FOCUSS) \cite{saxena2023sparse}, and its performance is assessed relative to Bayesian Cramer-Rao lower bound (BCRLB) references. Table \ref{SoTA} highlights the key differences between the proposed method and representative prior studies, and the subsequent subsection details the principal contributions.

\vspace{-4mm}
\subsection{Contributions}
\begin{enumerate}
\item An AP-STS signaling framework is formulated for CP-aided MIMO DCO-OTFS VLC system under arbitrary transmit and receive pulse-shaping filters. Owing to the orthogonality of the affine precoders, the receiver can separate the pilot and data contributions via post-multiplication with the corresponding precoder matrices, thereby suppressing pilot-data interference.

\item  A pilot-aided VBI (PA-VBI) approach is developed for MIMO DCO-OTFS, wherein DD-domain sparsity is leveraged to enhance CSI estimation accuracy. The resulting formulation is cast within a pilot-aided BL model, in which the sparse coefficients are governed by a hyper-parameter, yielding a hierarchical Bayesian structure that infers the full posterior distribution of the unknown quantities from the observations rather than producing only a point estimate. The pilot-aided BL objective is addressed via PA-VBI by selecting an approximating distribution from a constrained family and optimizing it to reduce the Kullback-Leibler (KL) divergence with respect to the target posterior, which provides computational tractability and convergence to a local optimum.

\item In addition, a data-aided VBI (DA-VBI) scheme for joint CSI estimation and data detection is proposed. By integrating a modified data-decision mechanism within the VBI framework, the approach jointly executes CE and symbol detection, yielding more accurate CSI than pilot-only methods. 

\item Closed-form BCRLB expressions are obtained to analytically characterize the performance of the proposed PA-VBI and DA-VBI estimators. The anticipated gains are then substantiated under diverse operating conditions using NMSE, pilot length, and SER as the principal assessment metrics.

\end{enumerate}
\vspace{-3mm}
\subsection{Organization}
The rest of this paper is structured as follows. Section II presents the data-aided AP-STS MIMO DCO-OTFS VLC system model. Section III formulates sparse DD-domain CIR estimation for the proposed framework. Section IV describes the PA-VBI estimator and the DA-VBI joint CSI estimation and data detection scheme, along with the corresponding BCRLB analysis. Section VI discusses simulation results, and Section VII concludes the paper.

\textit{Notations:} 
The notation used in this work is summarized as follows. The operator $\mathrm{blkmtx}(\mathbf{A}_1,\mathbf{A}_2,\ldots,\mathbf{A}_N)$ denotes a block-diagonal matrix with diagonal blocks $\mathbf{A}_1,\mathbf{A}_2,\ldots,\mathbf{A}_N$, where the blocks may be rectangular. The superscripts $(\cdot)^{T}$, $(\cdot)^{H}$, $(\cdot)^{*}$, and $(\cdot)^{-1}$ indicate transpose, Hermitian transpose, complex conjugation, and matrix inversion, respectively. The symbols $\otimes$ and $\mathrm{Tr}(\cdot)$ represent the Kronecker product and trace. Norms $||\cdot||_2$ and $||\cdot||_{F}$ denote the Euclidean and Frobenius norms, and $\mathbb{E}\{\cdot\}$ denotes expectation. Bold lowercase and uppercase letters denote vectors and matrices. The operators $\mathrm{vec}(\mathbf{A})$ and $\mathrm{vec}^{-1}(\mathbf{a})$ vectorize and reshape matrices, and we frequently use $\mathrm{vec}(\mathbf{A}\mathbf{B}\mathbf{C})=(\mathbf{C}^{T}\otimes \mathbf{A})\mathrm{vec}(\mathbf{B})$.

\vspace{-3mm}
\section{Data-Aided AP-STS MIMO DCO-OTFS VLC System Model}
Consider an AP-STS MIMO DCO-OTFS VLC system with frame duration $T_d = NT$ and bandwidth $B = M\Delta f$, where $T$ and $\Delta f$ denote the symbol interval and subcarrier spacing, respectively, and satisfy $T\Delta f=1$. The TF grid comprises $N$ symbols along the time axis and $M$ subcarriers along the frequency axis. The associated DD grid is sampled with resolutions $\Delta \nu = \frac{1}{T_d}$ and $\Delta \tau = \frac{1}{B}$. The transceiver employs $N_t$ LEDs and $N_r$ PDs.

\vspace{-5mm}

\subsection{Data-Aided AP-STS MIMO DCO-OTFS Modulation}
For the $t$th LED, $1\le t\le N_t$, assume that $N$ is even and define the set of independent Doppler indices as
$\mathcal{K}_{\mathrm{a}}=\left\{1,2,\ldots,\frac{N}{2}-1\right\}$, with cardinality $N_{\mathrm{a}}=\left|\mathcal{K}_{\mathrm{a}}\right|=\frac{N}{2}-1$.
Let the data and pilot symbol matrices be $\mathbf{S}_{t}^{d}\in\mathbb{C}^{M\times K_1}$ and $\mathbf{S}_{t}^{p}\in\mathbb{C}^{M\times K_2}$, respectively, where typically $K_1+K_2=N_{\mathrm{a}}$.
The entries of $\mathbf{S}_{t}^{d}$ and $\mathbf{S}_{t}^{p}$ are selected from an appropriate constellation with average powers $\sigma_d^2$ and $\sigma_p^2$, respectively, satisfying
$
\mathbb {E}\left\{ \mathbf {S}_{t}^{d}(\mathbf {{S}}^{d}_{t})^{H}\right\} =\sigma _{d}^{2} K_1 \mathbf {I}_{M},
\qquad
\mathrm {Tr}\left (\mathbf {S}_{t}^{p}(\mathbf {S}_{t}^{p})^{H}\right) = \sigma _{p}^{2} MK_{2},
$
and $\sigma _{d}^{2}+\sigma _{p}^{2}=\frac{1}{2}$.
In the proposed AP-STS construction, the data and pilot blocks are affine-precoded across the independent Doppler bins through semi-orthogonal transmit precoder matrices $\mathbf{D}\in\mathbb{C}^{N_{\mathrm{a}}\times K_1}$ and $\mathbf{P}\in\mathbb{C}^{N_{\mathrm{a}}\times K_2}$ that obey
\vspace{-3mm}
\begin{align}\label{eq:semi_orth}
\mathbf {P}^{H}\mathbf {P}= \mathbf {I}_{K_{2}},\mathbf {D}^{H} \mathbf {D}=\mathbf {I}_{K_{1}}, 
\mathbf {P}^{H}\mathbf {D} =\mathbf {0}_{K_{2}},\mathbf {D}^{H}\mathbf {P}=\mathbf {0}_{K_{1}}.
\end{align}
Such matrices can be obtained directly from any unitary matrix $\mathbf {U}\in \mathbb {C}^{N_{\mathrm{a}} \times N_{\mathrm{a}}}$ by selecting its columns as
$\mathbf {D}=\mathbf {U}(:,1:K_{1})$ and $\mathbf {P}=\mathbf {U}(:,K_{1}+1:N_{\mathrm{a}})$.
Accordingly, the DD-domain AP-STS block for the $t$th LED over the active Doppler set is given by
\begin{equation}\label{eq:ap_sip_dd_active}
\mathbf{S}_{t,\mathrm{a}}^{\mathrm{dd}}=\mathbf{S}_{t}^{d}\mathbf{D}^{H}+\mathbf{S}_{t}^{p}\mathbf{P}^{H}\in\mathbb{C}^{M\times N_{\mathrm{a}}}.
\end{equation}
To generate a real-valued time-domain (TD) waveform required by DCO-OTFS signaling, Doppler-axis Hermitian symmetry is imposed on the full $N$-length DD grid using $\mathbf{S}_{t,\mathrm{a}}^{\mathrm{dd}}$.
Specifically, for each delay index $l\in\{0,\ldots,M-1\}$, define $\mathbf{S}_{t}^{\mathrm{dd}}\in\mathbb{C}^{M\times N}$ as \cite{xu2023optical,chen2025optical,zheng2021dco}
\begin{equation}\label{eq:hermitian}
\mathbf{S}_{t}^{\mathrm{dd}}(l,k)=
\begin{cases}
\mathbf{S}_{t,\mathrm{a}}^{\mathrm{dd}}(l,k), & k=1,2,\ldots,\frac{N}{2}-1,\\[2pt]
\big(\mathbf{S}_{t,\mathrm{a}}^{\mathrm{dd}}(l,N-k)\big)^{*}, & k=\frac{N}{2}+1,\ldots,N-1,\\[2pt]
0, & k=0,\frac{N}{2},
\end{cases}
\end{equation}
which enforces the desired Hermitian structure along the Doppler dimension.
The DD symbols are then transformed to the time-frequency (TF)-domain using the inverse symplectic finite Fourier transform (ISFFT) \cite{xu2023optical}
\begin{align}\label{eq:isfft}
\mathbf{S}_{t}^{\mathrm{tf}}(m,n)
= \frac{1}{\sqrt{NM}}\sum_{l=0}^{M-1}\sum_{k=0}^{N-1}\mathbf{S}_{t}^{\mathrm{dd}}(l,k)
e^{j2\pi\left(\frac{nk}{N}-\frac{ml}{M}\right)},
\end{align}
where $\mathbf{S}_{t}^{\mathrm{tf}}\in\mathbb{C}^{M\times N}$ denotes the TF-domain symbol matrix.
Equivalently,
\begin{equation}\label{eq:isfft_mat}
\mathbf{S}_{t}^{\mathrm{tf}}=\mathbf{F}_M \mathbf{S}_{t}^{\mathrm{dd}}\mathbf{F}_N^{H},
\end{equation}
with $\mathbf{F}_M$ and $\mathbf{F}_N$ representing the unitary discrete Fourier transform (DFT) matrices of dimensions $M$ and $N$, respectively.
Let $p_{\mathrm{tx}}(t)$ denote the transmit pulse of duration $T$.
The corresponding TD symbol matrix follows from the Heisenberg transform as \cite{xu2023optical,cai2025power,zheng2021dco}
\begin{align}\label{eq:tx_mat}
\mathbf{X}_{t}=\mathbf{P}_{\mathrm{tx}}\mathbf{F}_M^{H}\mathbf{S}_{t}^{\mathrm{tf}}
=\mathbf{P}_{\mathrm{tx}}\mathbf{S}_{t}^{\mathrm{dd}}\mathbf{F}_N^{H},
\end{align}
where $\mathbf{P}_{\mathrm{tx}}=\mathrm{diag}\{p_{\mathrm{tx}}(\frac{pT}{M})\}_{p=0}^{M-1}\in\mathbb{R}^{M\times M}$.
Under \eqref{eq:hermitian}, the Doppler-axis inverse DFT (IDFT) produces a real-valued TD waveform.
In particular, letting $\tilde{\omega}_{N}=e^{j\frac{2\pi}{N}}$ and denoting the $(l,n)$th entry of $\mathbf{S}_{t}^{\mathrm{dd}}\mathbf{F}_N^{H}$ by $\tilde{x}_t[l,n]$, we obtain \cite{xu2023optical,cai2025power,zheng2021dco}
\begin{align}\label{eq:hermitian_property}
\tilde{x}_t[l,n]
&=\frac{1}{\sqrt{N}}\sum_{k=0}^{N-1}\mathbf{S}_{t}^{\mathrm{dd}}(l,k)\tilde{\omega}_{N}^{nk} \nonumber \\
&=\frac{1}{\sqrt{N}}\sum_{k=1}^{\frac{N}{2}-1}\Big(\mathbf{S}_{t}^{\mathrm{dd}}(l,k)\tilde{\omega}_{N}^{nk}+\mathbf{S}_{t}^{\mathrm{dd}}(l,N-k)\tilde{\omega}_{N}^{n(N-k)}\Big) \nonumber \\
&=\frac{1}{\sqrt{N}}\sum_{k=1}^{\frac{N}{2}-1}\Big(\mathbf{S}_{t}^{\mathrm{dd}}(l,k)\tilde{\omega}_{N}^{nk}+\mathbf{S}_{t}^{\mathrm{dd}}(l,k)^{*}\tilde{\omega}_{N}^{-nk}\Big) \nonumber \\
&=\frac{2}{\sqrt{N}}\Re\left\{\sum_{k=1}^{\frac{N}{2}-1}\mathbf{S}_{t}^{\mathrm{dd}}(l,k)\tilde{\omega}_{N}^{nk}\right\},
\end{align}
which implies $\mathbf{S}_{t}^{\mathrm{dd}}\mathbf{F}_N^{H}\in\mathbb{R}^{M\times N}$ and hence $\mathbf{X}_t\in\mathbb{R}^{M\times N}$ due to the real diagonal weighting $\mathbf{P}_{\mathrm{tx}}$.
Thereafter, each column of $\mathbf{X}_t$ is transmitted independently via parallel-to-serial (P/S) conversion, followed by the insertion of a cyclic prefix (CP) of length $L$ per column.
For DCO-OTFS signaling, an appropriate direct current (DC) bias is added to ensure nonnegative optical intensity; the corresponding biasing and clipping operations are omitted for brevity.
The next subsection presents the DD-domain VLC channel model under the adopted limited Doppler support (LDS) assumption.
\vspace{-3mm}
\subsection{DD-Domain VLC Channel Model with Limited Doppler Support (LDS)}
Let $h_{r,t}(\tau,\nu)$ represent the DD-domain VLC channel between the $t$th LED and the $r$th PD, where $1 \le r \le N_r$. Since practical indoor VLC environments typically contain only a small number of dominant reflectors, the DD-domain channel can be modeled as a sparse superposition of discrete propagation paths. Moreover, because mobility-induced Doppler spread is generally limited in indoor scenarios, the Doppler indices are assumed to lie within a limited Doppler support (LDS) set of size $N_{\nu} \ll N$, defined as $\mathcal{Q} \triangleq \left\{-\frac{N_{\nu}-1}{2}, -\frac{N_{\nu}-1}{2}+1, \ldots, \frac{N_{\nu}-1}{2}\right\}.$
Accordingly, the DD-domain channel is expressed as \cite{hadani2017orthogonal,raviteja2019otfs,sharma2023hyperparameter}
\begin{equation}\label{eq:dd_channel_mimo}
h_{r,t}(\tau,\nu)=\sum_{i=1}^{L_p} h_{i,r,t}\,\delta(\tau-\tau_i)\,\delta(\nu-\nu_i),
\end{equation}
where $h_{i,r,t}\in\mathbb{R}_{+}$ denotes the gain of the $i$th path, $\tau_i$ and $\nu_i$ are the associated delay and Doppler shifts, $L_p$ is the number of dominant multipath components, and $\delta(\cdot)$ denotes the Dirac delta function. Following the standard underspread OTFS channel model \cite{hadani2017orthogonal,raviteja2019otfs}, the path delays are assumed to be integer multiples of the delay resolution, that is, $\tau_i=l_i\Delta\tau$ with $\Delta\tau=\frac{1}{M\Delta f}$. In contrast, the Doppler shifts may be fractional with respect to the Doppler resolution $\Delta\nu=\frac{1}{NT}$. Specifically, the Doppler shift of the $i$th path is written as $\nu_i=\frac{k_i}{NT}=k_i\Delta\nu$, where $k_i=\mathrm{round}(k_i)+\kappa_i$ and $|\kappa_i|<0.5$. Here, $\mathrm{round}(k_i)\in\mathbb{Z}$ denotes the nearest integer Doppler-bin index, while $\kappa_i$ accounts for the fractional Doppler offset. Under the LDS assumption, the integer part of the Doppler index satisfies $\mathrm{round}(k_i)\in\mathcal{Q}$. Furthermore, for a typical underspread channel, the delay and Doppler indices satisfy $l_i \ll M$ and $|k_i| \ll N$ \cite{hadani2017orthogonal,raviteja2019otfs}. Since reflected optical components undergo stronger attenuation at larger delays, the power delay profile is modeled as an exponentially decaying function across the delay bins \cite{saxena2023sparse,liao2023sparse}. Accordingly, the magnitude of the $i$th tap is given by
\begin{equation}\label{eq:exp_decay_mimo}
h_{i,r,t}=
\frac{e^{-l_{i,r,t}\Delta\tau/\tau_{\mathrm{rms},r,t}}}
{\sum_{n=0}^{L_p-1} e^{-n\Delta\tau/\tau_{\mathrm{rms},r,t}}},
\end{equation}
where $l_{i,r,t}\in\{0,1,\ldots,L_p-1\}$ is the delay-bin index of the $i$th path and $\tau_{\mathrm{rms},r,t}$ denotes the root-mean-square (RMS) delay spread \cite{saxena2023sparse,liao2023sparse}. In general, $\tau_{\mathrm{rms},r,t}\in[0.5\Delta\tau,\,1.5\Delta\tau]$, and its exact value depends on the room geometry and the average reflectivity of the surrounding surfaces \cite{saxena2023sparse,liao2023sparse}.

\vspace{-4mm}
\subsection{Data-Aided AP-STS MIMO DCO-OTFS Demodulation}
Let $\mathbf{x}_{t,n}\in\mathbb{R}^{M\times 1}$, $0\le n\le N-1$, denote the $n$th column of the TD symbol matrix $\mathbf{X}_t$ transmitted by the $t$th LED, and let $\mathbf{r}_{r,n}\in\mathbb{C}^{M\times 1}$ denote the corresponding received sample vector at the $r$th PD. The $p$th entry of $\mathbf{r}_{r,n}$, denoted by $r_{r,n}(p)$ for $0\le p\le M-1$, is given by
\vspace{-3mm}
\begin{equation}\label{C1}
r_{r,n}(p)=\sum_{t=1}^{N_t}\sum_{i=1}^{L_p} h_{i,r,t} 
e^{j2\pi \frac{k_i(p-l_i)}{MN}} x_{t,n}\big([p-l_i]_M\big)+w_{r,n}(p),
\end{equation}
where $x_{t,n}(p)$ is the $p$th element of $\mathbf{x}_{t,n}$ and $w_{r,n}(p)$ denotes the noise sample.
Define the stacked vectors $\mathbf{r}_{r,n}=\big[r_{r,n}(0),r_{r,n}(1),\ldots,r_{r,n}(M-1)\big]^T\in\mathbb{C}^{M\times 1},$
and
$\mathbf{w}_{r,n}=\big[w_{r,n}(0),w_{r,n}(1),\ldots,w_{r,n}(M-1)\big]^T\in\mathbb{C}^{M\times 1}.$
Using \eqref{C1}, the receive vector can be written compactly as
\vspace{-3mm}
\begin{align}
\mathbf{r}_{r,n}
&=\sum_{t=1}^{N_t}\sum_{i=1}^{L_p} h_{i,r,t}
\left(\bar{\mathbf{\Pi}}\right)^{l_i}
\left(\bar{\mathbf{\Delta}}_{l_i,{k_i}}\right)\mathbf{x}_{t,n}
+\mathbf{w}_{r,n}\nonumber\\
&=\sum_{t=1}^{N_t}\bar{\mathbf{H}}_{r,t}\mathbf{x}_{t,n}+\mathbf{w}_{r,n},
\end{align}
where $\bar{\mathbf{H}}_{r,t}\in\mathbb{C}^{M\times M}$ is defined as
$
\bar{\mathbf{H}}_{r,t}\triangleq\sum_{i=1}^{L_p} h_{i,r,t}
\left(\bar{\mathbf{\Pi}}\right)^{l_i}\left(\bar{\mathbf{\Delta}}_{l_i,{k_i}}\right).
$
Here, $\bar{\mathbf{\Pi}}$ denotes an $M\times M$ permutation matrix and $\bar{\mathbf{\Delta}}_{l_i,k_i}\in\mathbb{C}^{M\times M}$ is the diagonal matrix 
\vspace{-3mm}
\begin{align}
\bar{\boldsymbol{\Delta}}_{l_i,k_i}=
\begin{cases}
\mathrm{diag}\left\{1,\omega,\ldots,\omega^{M-l_i-1},\omega^{-l_i},\ldots,\omega^{-1}\right\}, & \hspace{-3mm}l_i\neq 0,\\
\mathrm{diag}\left\{1,\omega,\ldots,\omega^{M-1}\right\}, & \hspace{-3mm}l_i=0,
\end{cases}
\end{align}
with $\omega=e^{j2\pi\frac{k_i}{MN}}$ \cite{srivastava2021bayesian}.
By concatenating $\mathbf{r}_{r,n}$ across $n$, define
$
\mathbf{R}_r\triangleq\big[\mathbf{r}_{r,0},\mathbf{r}_{r,1},\ldots,\mathbf{r}_{r,N-1}\big]\in\mathbb{C}^{M\times N}, $ $
\mathbf{W}_r\triangleq\big[\mathbf{w}_{r,0},\mathbf{w}_{r,1},\ldots,\mathbf{w}_{r,N-1}\big]\in\mathbb{C}^{M\times N}. $
Then,
\vspace{-2mm}
\begin{equation}
\mathbf{R}_r=\sum_{t=1}^{N_t}\bar{\mathbf{H}}_{r,t}\mathbf{X}_t+\mathbf{W}_r.
\end{equation}
Next, DCO-OTFS demodulation is applied to $\mathbf{R}_r$. The receiver first performs pulse shaping with $p_{\mathrm{rx}}(t)$ (duration $T$), and then applies an $M$-point fast Fourier transform (FFT) across each column to form the TF-domain output $\mathbf{Y}^{\mathrm{tf}}_r\in\mathbb{C}^{M\times N}$ given as 
\vspace{-2mm}
\begin{equation}
\mathbf{Y}^{\mathrm{tf}}_r=\mathbf{F}_M \mathbf{P}_{\mathrm{rx}} \mathbf{R}_r,
\end{equation}
where $\mathbf{P}_{\mathrm{rx}}=\mathrm{diag}\left\{p_{\mathrm{rx}}^{*}\left(\frac{pT}{M}\right)\right\}_{p=0}^{M-1}$.
Subsequently, the DD-domain observation $\mathbf{Y}^{\mathrm{dd}}_r\in\mathbb{C}^{M\times N}$ is obtained via the symplectic FFT (SFFT) along the Doppler axis
\vspace{-2mm}
\begin{equation}
\mathbf{Y}^{\mathrm{dd}}_r=\mathbf{F}_M^{H}\mathbf{Y}^{\mathrm{tf}}_r\mathbf{F}_N
=\mathbf{P}_{\mathrm{rx}} \mathbf{R}_r\mathbf{F}_N.
\end{equation}
Substituting $\mathbf{R}_r$, and then $\mathbf{X}_t$, yields the DD-domain input-output relation for the AP-STS MIMO DCO-OTFS model
\vspace{-2mm}
\begin{equation}\label{eq:8}
\mathbf{Y}^{\mathrm{dd}}_r=\sum_{t=1}^{N_t}\bar{\mathbf{H}}^{\mathrm{dd}}_{r,t}\mathbf{S}^{\mathrm{dd}}_t+\mathbf{W}^{\mathrm{dd}}_r,
\end{equation}
where $\bar{\mathbf{H}}^{\mathrm{dd}}_{r,t}=\mathbf{P}_{\mathrm{rx}}\bar{\mathbf{H}}_{r,t}\mathbf{P}_{\mathrm{tx}}\in\mathbb{C}^{M\times M}$ and can be written as
\vspace{-2mm}
\begin{equation}\label{eq:9}
\bar{\mathbf{H}}^{\mathrm{dd}}_{r,t}
=\sum_{i=1}^{L_p} h_{i,r,t} \mathbf{P}_{\mathrm{rx}}
\left(\bar{\mathbf{\Pi}}\right)^{l_i}
\left(\bar{\mathbf{\Delta}}_{l_i}\right)^{k_i}
\mathbf{P}_{\mathrm{tx}},
\end{equation}
and $\mathbf{W}^{\mathrm{dd}}_r=\mathbf{P}_{\mathrm{rx}}\mathbf{W}_r\mathbf{F}_N$.
Stacking the DD-domain outputs over all PDs gives $\mathbf{Y}^{\mathrm{dd}}
=\big[(\mathbf{Y}^{\mathrm{dd}}_1)^T (\mathbf{Y}^{\mathrm{dd}}_2)^T \cdots (\mathbf{Y}^{\mathrm{dd}}_{N_r})^T\big]^T\in\mathbb{C}^{MN_r\times N}$ 
such that
\vspace{-2mm}
\begin{equation}\label{eq:48}
\mathbf{Y}^{\mathrm{dd}}=\widetilde{\mathbf{H}}^{\mathrm{dd}}\mathbf{S}^{\mathrm{dd}}+\mathbf{W}^{\mathrm{dd}},
\end{equation}
with
$\mathbf{S}^{\mathrm{dd}}=\big[(\mathbf{S}^{\mathrm{dd}}_1)^T(\mathbf{S}^{\mathrm{dd}}_2)^T \cdots (\mathbf{S}^{\mathrm{dd}}_{N_t})^T\big]^T\in\mathbb{C}^{MN_t\times N},$
$\mathbf{W}^{\mathrm{dd}}=\big[(\mathbf{W}^{\mathrm{dd}}_1)^T (\mathbf{W}^{\mathrm{dd}}_2)^T \cdots (\mathbf{W}^{\mathrm{dd}}_{N_r})^T\big]^T\in\mathbb{C}^{MN_r\times N},$
and the DD-domain MIMO channel $\widetilde{\mathbf{H}}^{\mathrm{dd}}\in\mathbb{C}^{MN_r\times MN_t}$ given by
\vspace{-3mm}
\begin{align}
&\widetilde{\mathbf{H}}^{\mathrm{dd}}
=\mathrm{blkmtx}\left\{\bar{\mathbf{H}}^{\mathrm{dd}}_{r,t}\right\}_{r=1,t=1}^{N_r,N_t}\notag\\
&=(\mathbf{I}_{N_r}\otimes \mathbf{P}_{\mathrm{rx}})
\Big[\mathrm{blkmtx}\left\{\bar{\mathbf{H}}_{r,t}\right\}_{r=1,t=1}^{N_r,N_t}\Big]
(\mathbf{I}_{N_t}\otimes \mathbf{P}_{\mathrm{tx}}).
\label{eq:49}
\end{align}
Since AP-STS precoding is applied only over the independent Doppler bins, pilot-data separation is performed on the corresponding DD columns. Let $\mathbf{Y}_{r,\mathrm{a}}^{\mathrm{dd}}$ denote the restriction of $\mathbf{Y}_r^{\mathrm{dd}}$ to the independent Doppler indices, and define $\mathbf{W}_{r,\mathrm{a}}^{\mathrm{dd}}$ analogously. Using $\mathbf{S}_{t,\mathrm{a}}^{\mathrm{dd}}$, the restricted model becomes
\vspace{-2mm}
\begin{equation}\label{eq:Ydd_a}
\mathbf{Y}_{r,\mathrm{a}}^{\mathrm{dd}}
=\sum_{t=1}^{N_t}\bar{\mathbf{H}}^{\mathrm{dd}}_{r,t}\mathbf{S}_{t,\mathrm{a}}^{\mathrm{dd}}
+\mathbf{W}_{r,\mathrm{a}}^{\mathrm{dd}}.
\end{equation}
Substituting \eqref{eq:ap_sip_dd_active} into \eqref{eq:Ydd_a} yields
\vspace{-2mm}
\begin{equation}\label{eq:Ydd_a_expand}
\mathbf{Y}_{r,\mathrm{a}}^{\mathrm{dd}}
=\sum_{t=1}^{N_t}\bar{\mathbf{H}}^{\mathrm{dd}}_{r,t}\left(\mathbf{S}_t^{d}\mathbf{D}^{H}+\mathbf{S}_t^{p}\mathbf{P}^{H}\right)
+\mathbf{W}_{r,\mathrm{a}}^{\mathrm{dd}}.
\end{equation}
By exploiting the property of \eqref{eq:semi_orth}, the decoupled data observation is given by post-multiplication with $\mathbf{D}$
\vspace{-2mm}
\begin{align}
\mathbf{Y}_{r}^{\mathrm{dd},d}
=\mathbf{Y}_{r,\mathrm{a}}^{\mathrm{dd}}\mathbf{D}&=\sum_{t=1}^{N_t}\bar{\mathbf{H}}^{\mathrm{dd}}_{r,t}
\left(\mathbf{S}_t^{d}\mathbf{D}^{H}\mathbf{D}+\mathbf{S}_t^{p}\mathbf{P}^{H}\mathbf{D}\right)
+\mathbf{W}_{r,\mathrm{a}}^{\mathrm{dd}}\mathbf{D}\notag\\
&=\sum_{t=1}^{N_t}\bar{\mathbf{H}}^{\mathrm{dd}}_{r,t}\mathbf{S}_t^{d}
+\mathbf{W}_{r}^{\mathrm{dd},d},
\label{eq:51}
\end{align}
where $\mathbf{W}_{r}^{\mathrm{dd},d}\triangleq\mathbf{W}_{r,\mathrm{a}}^{\mathrm{dd}}\mathbf{D}$, and the last step follows from \eqref{eq:semi_orth}. Finally, stacking $\mathbf{Y}_{r}^{\mathrm{dd},d}$ across all PDs yields $
\widetilde{\mathbf{Y}}^{\mathrm{dd},d}
=\left[(\mathbf{Y}_{1}^{\mathrm{dd},d})^{T} (\mathbf{Y}_{2}^{\mathrm{dd},d})^{T} \cdots (\mathbf{Y}_{N_r}^{\mathrm{dd},d})^{T}\right]^{T} \in\mathbb{C}^{MN_r\times K_1}.$
The resulting decoupled data model is
\vspace{-2mm}
\begin{equation}\label{eq:53}
\widetilde{\mathbf{Y}}^{\mathrm{dd},d}
=\widetilde{\mathbf{H}}^{\mathrm{dd}}\widetilde{\mathbf{S}}^{d}
+\widetilde{\mathbf{W}}^{\mathrm{dd},d},
\end{equation}
where
$
\widetilde{\mathbf{S}}^{d}=\left[(\mathbf{S}_{1}^{d})^{T}(\mathbf{S}_{2}^{d})^{T}\cdots(\mathbf{S}_{N_t}^{d})^{T}\right]^{T}\in\mathbb{C}^{MN_t\times K_1},$
$\widetilde{\mathbf{W}}^{\mathrm{dd},d}
=\left[(\mathbf{W}_{1}^{\mathrm{dd},d})^{T}(\mathbf{W}_{2}^{\mathrm{dd},d})^{T}\cdots(\mathbf{W}_{N_r}^{\mathrm{dd},d})^{T}\right]^{T}\in\mathbb{C}^{MN_r\times K_1}.$
An LMMSE detector for $\widetilde{\mathbf{S}}^{d}$ is given by
\vspace{-3mm}
\begin{align}
\widetilde{\mathbf{S}}^{d}_{\mathrm{LMMSE}}&=\left((\widetilde{\mathbf{H}}^{\mathrm{dd}})^{H}\widetilde{\mathbf{R}}_{{w}_{d}}^{-1}\widetilde{\mathbf{H}}^{\mathrm{dd}}
+\frac{1}{\sigma_d^{2}}\mathbf{I}_{MN_t}\right)^{-1} \nonumber \\
&\times (\widetilde{\mathbf{H}}^{\mathrm{dd}})^{H}\widetilde{\mathbf{R}}_{{w}_{d}}^{-1}
\widetilde{\mathbf{Y}}^{\mathrm{dd},d},
\label{eq:54}
\end{align}
where $\widetilde{\mathbf{R}}_{w_d}=(\mathbf{I}_{N_r}\otimes\mathbf{R}_{w_d})\in\mathbb{C}^{MN_r\times MN_r}$ denotes the noise covariance matrix and $\mathbf{R}_{w_d} = \beta^{-1}\mathbf{I}_M$ with $\beta$ denote the noise precision parameter. The detected symbols are subsequently demodulated using nearest-neighbor decoding for the chosen transmit constellation.
\vspace{-3mm}

\section{Sparse DD-domain CIR estimation model for AP-STS MIMO DCO-OTFS VLC Systems}
This section formulates a sparse CIR estimation model for the AP-STS MIMO DCO-OTFS VLC system under the LDS assumption. Since AP-STS precoding is applied only across the independent Doppler bins, pilot-aided CSI estimation is carried out using the DD-domain output restricted to these indices. Let $\mathbf{Y}_{r,\mathrm{a}}^{\mathrm{dd}}\in\mathbb{C}^{M\times N_{\mathrm{a}}}$ denote the DD-domain received matrix at the $r$th PD over the independent Doppler bins, and define $\mathbf{W}_{r,\mathrm{a}}^{\mathrm{dd}}\in\mathbb{C}^{M\times N_{\mathrm{a}}}$ similarly. For CSI acquisition, the pilot observation $\mathbf{Y}^{\mathrm{dd},p}_r\in\mathbb{C}^{M\times K_{2}}$ is separated from the data component by post-multiplying $\mathbf{Y}_{r,\mathrm{a}}^{\mathrm{dd}}$ with the transmit precoder (TPC) matrix $\mathbf{P}$, that is, $\mathbf{Y}^{\mathrm{dd},p}_r=\mathbf{Y}_{r,\mathrm{a}}^{\mathrm{dd}}\mathbf{P}$. Using the DD-domain input-output relation and applying \eqref{eq:semi_orth}, we obtain
\vspace{-5mm}
\begin{align}
\mathbf{Y}_{r}^{\mathrm{dd},p}
=\mathbf{Y}_{r,\mathrm{a}}^{\mathrm{dd}}\mathbf{P}
&=\sum_{t=1}^{N_{t}}
\bar{\mathbf{H}}_{r,t}^{\mathrm{dd}}
\big(
\mathbf{S}_{t}^{d}\mathbf{D}^{H}\mathbf{P}
+\mathbf{S}_{t}^{p}\mathbf{P}^{H}\mathbf{P}
\big)
+\mathbf{W}_{r,\mathrm{a}}^{\mathrm{dd}}
\mathbf{P} \nonumber\\
&=\sum_{t=1}^{N_{t}}\bar{\mathbf{H}}_{r,t}^{\mathrm{dd}}\mathbf{S}_{t}^{p}
+\mathbf{W}_{r}^{\mathrm{dd},p},
\label{eq:55}
\end{align}
where $\mathbf{W}_{r}^{\mathrm{dd},p}\triangleq \mathbf{W}_{r,\mathrm{a}}^{\mathrm{dd}}\mathbf{P}\in\mathbb{C}^{M\times K_2}$. For the DD-domain MIMO DCO-OTFS VLC channel, let the maximum delay spread be $M_{\tau}$, while the Doppler support is restricted to the set $\mathcal{Q}$, where $N_{\nu}=|\mathcal{Q}|\ll N$. Under the underspread channel assumption, the channel indices satisfy $l_{\max}=\max(l_i)<M_{\tau}\ll M$ and $k_{\max}=\max(k_j)\leq \frac{N_{\nu}-1}{2}\ll N$. To account for fractional Doppler, a refined Doppler grid of size $M_{\tau}\times G_{\nu}$ is introduced, with $N_{\nu}=|\mathcal{Q}|\ll G_{\nu}$. In this model, the $j$th Doppler-grid point, for $0\leq j\leq G_{\nu}-1$, corresponds to the Doppler shift $\nu_j=\frac{k_j}{NT}$ Hz, where $k_j=-\frac{N_{\nu}-1}{2}+\frac{j(N_{\nu}-1)}{G_{\nu}-1}.$
Let $h_{i,j,r,t}$ denote the path gain associated with the $i$th delay tap and the $j$th Doppler tap for the link between the $r$th PD and the $t$th LED, given as
\vspace{-2mm}
\begin{equation}
h_{r,t}(\tau,\nu)
=
\sum_{i=0}^{M_{\tau}-1}\sum_{j=0}^{G_{\nu}-1}
h_{i,j,r,t}
\delta(\tau-\tau_i)\,\delta(\nu-\nu_j).
\label{eq:56}
\end{equation} Substituting the expression of $\bar{\mathbf{H}}_{r,t}^{\mathrm{dd}}$ into \eqref{eq:55}$,$ followed by vectorization, yields
\vspace{-2mm}
\begin{align}
\mathbf{y}_{r}^{\mathrm{dd},p}
&=\sum_{t=1}^{N_{t}}
\mathrm{vec}\bigg(
\mathbf{P}_{\mathrm{rx}}
\bigg(
\sum_{i=0}^{M_{\tau}-1}\sum_{j=0}^{G_{\nu}-1}
h_{i,j,r,t}(\bar{\boldsymbol{\Pi}})^{i}\bar{\boldsymbol{\Delta}}_{i,j}
\bigg)
\mathbf{P}_{\mathrm{tx}}\mathbf{S}_{t}^{p}
\bigg) \nonumber \\
&+\mathbf{w}_{r}^{\mathrm{dd},p},
\label{eq:57}
\end{align}
where $\mathbf{y}_{r}^{\mathrm{dd},p}=\mathrm{vec}(\mathbf{Y}_{r,\mathrm{a}}^{\mathrm{dd}}\mathbf{P})\in\mathbb{C}^{MK_{2}\times 1}$ and
$\mathbf{w}_{r}^{\mathrm{dd},p}=\mathrm{vec}(\mathbf{W}_{r,\mathrm{a}}^{\mathrm{dd}}\mathbf{P})\in\mathbb{C}^{MK_{2}\times 1}$. Here, $\bar{\boldsymbol{\Delta}}_{i,j}\in\mathbb{C}^{M\times M}$ denotes the diagonal matrix given in \cite{srivastava2021bayesian} as
\vspace{-2mm}
\begin{align}
\bar{\boldsymbol{\Delta}}_{i,j}
=
\begin{cases}
\mathrm{diag}\left\{1,\bar{\omega}_j,\cdots,\bar{\omega}_j^{M-i-1},\bar{\omega}_j^{-i},\cdots,\bar{\omega}_j^{-1}\right\},&\text{\hspace{-3mm}if } i\neq 0,\\
\mathrm{diag}\left\{1,\bar{\omega}_j,\cdots,\bar{\omega}_j^{M-1}\right\},& \text{\hspace{-3mm}if } i=0,
\end{cases}
\end{align}
where $\bar{\omega}_j=e^{j2\pi \frac{k_j}{MN}}.$
After simplification, we have
\vspace{-2mm}
\begin{equation}
\mathbf{y}_{r}^{\mathrm{dd},p}
=
\sum_{t=1}^{N_{t}}\sum_{i=0}^{M_{\tau}-1}\sum_{j=0}^{G_{\nu}-1}
\boldsymbol{\omega}_{i,j,t}^{p}h_{i,j,r,t}
+\mathbf{w}_{r}^{\mathrm{dd},p},
\label{eq:58}
\end{equation}
where
$
\boldsymbol{\omega}_{i,j,t}^{p}
=
\big(\mathbf{I}_{K_{2}}\otimes \mathbf{P}_{\mathrm{rx}}(\bar{\boldsymbol{\Pi}})^{i}(\bar{\boldsymbol{\Delta}}_{i,j})\mathbf{P}_{\mathrm{tx}}\big)\mathbf{s}_{t}^{p}
\in\mathbb{C}^{MK_{2}\times 1},
\qquad
\mathbf{s}_{t}^{p}=\mathrm{vec}(\mathbf{S}_{t}^{p})\in\mathbb{C}^{MK_{2}\times 1}.
$
Equivalently, \eqref{eq:58} admits the compact form
\vspace{-3mm}
\begin{equation}
\mathbf{y}_{r}^{\mathrm{dd},p}
=
\sum_{t=1}^{N_{t}}\boldsymbol{\Omega}_{t}^{p}\mathbf{h}_{r,t}
+\mathbf{w}_{r}^{\mathrm{dd},p},
\label{eq:59}
\end{equation}
where $\boldsymbol{\Omega}_{t}^{p}\in\mathbb{C}^{MK_{2}\times M_{\tau}G_{\nu}}$ is the dictionary matrix for the $t$th LED, defined as
$
\boldsymbol{\Omega}_{t}^{p}=\big[
\boldsymbol{\omega}_{0,0,t}^{p} \ \cdots \ \boldsymbol{\omega}_{0,G_{\nu}-1,t}^{p} \ \cdots \
\boldsymbol{\omega}_{M_{\tau}-1,0,t}^{p} \ \cdots \ \boldsymbol{\omega}_{M_{\tau}-1,G_{\nu}-1,t}^{p}\big],$ and $\mathbf{h}_{r,t} = \big[h_{0,0,r,t} \cdots h_{0,G_{\nu}-1,r,t} \cdots h_{M_{\tau}-1,0,r,t} \cdots$ $h_{M_{\tau}-1,G_{\nu}-1,r,t}
\big]^{T}\in\mathbb{R}_{+}^{M_{\tau}G_{\nu}\times 1}$ is the CIR vector. By stacking the LED-specific dictionaries, such that $\widetilde{\boldsymbol{\Omega}}^{p}=[\boldsymbol{\Omega}_{1}^{p},\boldsymbol{\Omega}_{2}^{p},\ldots,\boldsymbol{\Omega}_{N_{t}}^{p}]\in\mathbb{C}^{MK_{2}\times M_{\tau}G_{\nu}N_{t}}$, \eqref{eq:59} becomes
\vspace{-2mm}
\begin{equation}
\mathbf{y}_{r}^{\mathrm{dd},p}=
\widetilde{\boldsymbol{\Omega}}^{p}\mathbf{h}_{r}
+\mathbf{w}_{r}^{\mathrm{dd},p},
\label{eq:61}
\end{equation}
where
$\mathbf{h}_{r}=$ $[\mathbf{h}_{r,1}^{T},\mathbf{h}_{r,2}^{T},\ldots,\mathbf{h}_{r,N_{t}}^{T}]^{T}\in\mathbb{R}_{+}^{M_{\tau}G_{\nu}N_{t}\times 1}$.
Concatenating \eqref{eq:61} across all PDs yields the global observation model
\vspace{-4mm}
\begin{equation}
{\mathbf{y}}^{\mathrm{dd},p}
=
{\boldsymbol{\Omega}}^{p}{\mathbf{h}}
+{\mathbf{w}}^{p},
\label{eq:62}
\end{equation}
where
$
{\mathbf{y}}^{\mathrm{dd},p}=[(\mathbf{y}_{1}^{\mathrm{dd},p})^{T},(\mathbf{y}_{2}^{\mathrm{dd},p})^T,\ldots,(\mathbf{y}_{N_{r}}^{\mathrm{dd},p})^T]^T\in \mathbb{C}^{MK_2N_r\times 1},$
$
{\mathbf{h}}=[\mathbf{h}_{1}^T,\mathbf{h}_{2}^T,\ldots,(\mathbf{h}_{N_{r}})^T]^T\in\mathbb{R}_{+}^{M_{\tau}G_{\nu}N_{t}N_r\times 1},$ $
{\boldsymbol{\Omega}}^p = \mathbf{I}_{N_r} \otimes \widetilde{\boldsymbol{\Omega}}^p \in \mathbb{C}^{MK_2N_r \times M_\tau G_\nu N_t N_r},$
and
${\mathbf{w}}^{p}=[(\mathbf{w}_{1}^{\mathrm{dd},p})^T,(\mathbf{w}_{2}^{\mathrm{dd},p})^T,\ldots,(\mathbf{w}_{N_{r}}^{\mathrm{dd},p})^T]
\in\mathbb{C}^{MK_{2}N_{r}\times 1}$.
For \eqref{eq:62}, the conventional LMMSE estimate $\widehat{\mathbf{h}}_{\mathrm{LMMSE}}$ is
\vspace{-2mm}
\begin{equation}
\widehat{\mathbf{h}}_{\mathrm{LMMSE}}
=
\left(({\boldsymbol{\Omega}}^{p})^{H}\mathbf{R}_{w_{p}}^{-1}{\boldsymbol{\Omega}}^{p}+\mathbf{R}_{h}^{-1}\right)^{-1}
({\boldsymbol{\Omega}}^{p})^{H}\mathbf{R}_{w_{p}}^{-1}{\mathbf{y}}^{\mathrm{dd},p},
\label{eq:63}
\end{equation}
where $\mathbf{R}_{h}=\mathbb{E}[\mathbf{h}{\mathbf{h}}^{H}]$ is generally unavailable and is therefore set to $\mathbf{I}_{M_{\tau}G_{\nu}N_{t}N_r}$, and the noise covariance is given by
$\mathbf{R}_{w_{p}}=\beta^{-1}\big(\mathbf{I}_{N_r}\otimes\mathbf{I}_{K_{2}}\otimes\mathbf{P}_{\mathrm{rx}}\mathbf{P}_{\mathrm{rx}}^{H}\big)
\in\mathbb{C}^{MK_{2}N_r\times MK_{2}N_r}.$
A key limitation of \eqref{eq:63} is that it does not utilize the sparsity inherent to the DD-domain channel, which can otherwise be exploited to achieve substantially improved estimation accuracy. Motivated by this observation, the next subsection develops a pilot-aided VBI (PA-VBI) procedure for sparse CSI estimation in the  MIMO DCO-OTFS VLC system.
\vspace{-2mm}
\section{Pilot-aided VBI (PA-VBI) for AP-STS MIMO DCO-OTFS VLC systems}
In this section, a pilot-aided VBI (PA-VBI) procedure is utilized to estimate the sparse VLC DD-domain CIR vector ${\mathbf{h}}$ from the pilot observation ${\mathbf{y}}^{\mathrm{dd},p}$. In principle, this requires evaluating the marginal posterior conditioned on ${\mathbf{y}}^{\mathrm{dd},p}$, which involves a high-dimensional integral that is generally intractable \cite{liu2020uplink}. To circumvent this difficulty, we employ the VBI methodology to obtain a tractable approximation of the desired posterior.

Within the proposed BL model, the DD-domain CIR vector ${\mathbf{h}}$ is assigned a parameterized Gaussian prior $f(\mathbf{h}; \pmb{\Gamma}^{-1})$, where $\pmb{\Gamma}=\mathrm{diag}(\pmb{\gamma})$ and $\pmb{\gamma}=[\gamma_{1},\gamma_{2},\dots,\gamma_{M_{\tau}G_{\nu}N_{t}N_{r}}]^{T}\in\mathbb{R}^{M_{\tau}G_{\nu}N_{t}N_{r}\times 1}$ denotes a non-negative precision vector. Specifically, the prior factorizes as 
\vspace{-2mm}
\begin{align}\label{eq:h_prior_nsbl_otfs}
f({\mathbf{h}}|\pmb{\Gamma})
&=\prod_{l=1}^{M_{\tau}G_{\nu}N_{t}N_{r}}
\left(\frac{\gamma_l}{2\pi}\right)^{\frac{1}{2}}
\exp\left(-\frac{\gamma_l|h(l)|^{2}}{2}\right),
\end{align}
where $\gamma_l$ is the unknown precision hyperparameter associated with the $l$th element of ${\mathbf{h}}$, for $1\le l\le M_{\tau}G_{\nu}N_{t}N_{r}$ \cite{wipf2007empirical,liu2020uplink}. The hyperparameters are modeled using a Gamma prior
\vspace{-2mm}
\begin{align}\label{eq:gamma_prior_nsbl_otfs}
f(\pmb{\gamma})
&=\prod_{l=1}^{M_{\tau}G_{\nu}N_{t}N_{r}}
\mathrm{Gamma}(\gamma_{l};a,b) \nonumber \\
&=\prod_{l=1}^{M_{\tau}G_{\nu}N_{t}N_{r}}
\frac{b^{a}}{\Gamma(a)}\gamma_{l}^{a-1}\exp(-b\gamma_{l}),
\end{align}
where $\gamma_{l}>0$, and $a$ and $b$ are the shape and inverse-scale parameters, respectively. The Gamma function is defined as $\Gamma(a)=\int_{0}^{\infty} t^{a-1}e^{-t}dt$. This construction yields a conjugate prior for $f(h_l|\gamma_l)$, thereby enabling closed-form variational updates. Moreover, marginalizing $\gamma_l$ in $f(h_l)=\int f(h_l|\gamma_l)f(\gamma_l)d\gamma_l$ results in a Student-t type marginal for $h_l$, which naturally promotes sparsity in ${\mathbf{h}}$ \cite{wipf2007empirical,liu2020uplink}.

The observation noise ${\mathbf{w}}^{p}$ is modeled as circularly symmetric complex Gaussian, ${\mathbf{w}}^{p}\sim \mathcal{CN}(\mathbf{0},\mathbf{R}_{w_{p}})$, with covariance matrix $\mathbf{R}_{w_{p}}$. 
A Gamma prior is assigned to $\beta$ as $\beta\sim \mathrm{Gamma}(c,d),$
where $c$ and $d$ are the shape and inverse-scale parameters. Under these assumptions, the likelihood of the received pilot vector ${\mathbf{y}}^{\mathrm{dd},p}$ is
\vspace{-2mm}
\begin{align}\label{eq:likelihood_nsbl_otfs}
&f({\mathbf{y}}^{\mathrm{dd},p}|{\mathbf{h}},\beta) =\left(\frac{\beta}{\pi}\right)^{MK_{2}N_{r}} \nonumber \\ 
& \times
\exp\left(-\beta\big({\mathbf{y}}^{\mathrm{dd},p}-{\boldsymbol{\Omega}}^{p}{\mathbf{h}}\big)^{H}
\big({\mathbf{y}}^{\mathrm{dd},p}-{\boldsymbol{\Omega}}^{p}{\mathbf{h}}\big)\right).
\end{align}
Let the latent-variable set be $\pmb{\Psi}=\{\pmb{\gamma},{\mathbf{h}},\beta\}$ and the observable be ${\mathbf{y}}^{\mathrm{dd},p}$. The resulting hierarchical model implies the joint density factorization
\vspace{-2mm}
\begin{equation}\label{eq:joint_factorization_nsbl_otfs}
f({\mathbf{y}}^{\mathrm{dd},p},\pmb{\Psi})
=f({\mathbf{y}}^{\mathrm{dd},p}|{\mathbf{h}},\beta)f({\mathbf{h}}|\pmb{\gamma})f(\pmb{\gamma})f(\beta).
\end{equation}
Computing the maximum \textit{a posteriori} (MAP) characterization of $f(\pmb{\Psi}|{\mathbf{y}}^{\mathrm{dd},p})$ typically requires evaluating high-dimensional integrals. Hence, PA-VBI is adopted to construct an iterative approximation based on alternate updates \cite{wipf2007empirical,liu2020uplink}. To this end, an approximating density $q(\pmb{\Psi})$ is introduced, and its mismatch with the true posterior is quantified via the Kullback-Leibler (KL) divergence \cite{wipf2007empirical,liu2020uplink}
\vspace{-2mm}
\begin{equation}\label{eq:kl_nsbl_otfs}
\mathrm{KL}\left(q(\pmb{\Psi})\Vert f(\pmb{\Psi}|{\mathbf{y}}^{\mathrm{dd},p})\right)
=-\int q(\pmb{\Psi})\ln\frac{f(\pmb{\Psi}|{\mathbf{y}}^{\mathrm{dd},p})}{q(\pmb{\Psi})}d\pmb{\Psi}.
\end{equation}
Minimizing \eqref{eq:kl_nsbl_otfs} is equivalent to maximizing the evidence lower bound
\vspace{-3mm}
\begin{equation}\label{eq:q_opt_nsbl_otfs}
q^{*}(\pmb{\Psi})=\arg\max_{q(\pmb{\Psi})}\int q(\pmb{\Psi})\ln\frac{f({\mathbf{y}}^{\mathrm{dd},p},\pmb{\Psi})}{q(\pmb{\Psi})}d\pmb{\Psi}.
\end{equation}
Under the mean-field assumption \cite{liu2020uplink}, the variational density is factorized as
\vspace{-3mm}
\begin{equation}\label{eq:mf_factorization_nsbl_otfs}
q(\pmb{\Psi})=q(\pmb{\gamma})q({\mathbf{h}})q(\beta).
\end{equation}
Using the identity $\ln f({\mathbf{y}}^{\mathrm{dd},p})=\mathcal{L}(q(\pmb{\Psi}))+\mathrm{KL}\left(q(\pmb{\Psi})\Vert f(\pmb{\Psi}|{\mathbf{y}}^{\mathrm{dd},p})\right),$
the lower bound is
\vspace{-3mm}
\begin{equation}\label{eq:elbo_nsbl_otfs}
\mathcal{L}(q(\pmb{\Psi}))=\int q(\pmb{\Psi})\ln\frac{f({\mathbf{y}}^{\mathrm{dd},p},\pmb{\Psi})}{q(\pmb{\Psi})}d\pmb{\Psi}.
\end{equation}
Maximizing \eqref{eq:elbo_nsbl_otfs} with respect to each factor yields the update
\vspace{-3mm}
\begin{equation}\label{eq:q_update_general_nsbl_otfs}
q(\pmb{\Psi}_{t})\propto \exp\Big\{\mathbb{E}_{\sim q(\pmb{\Psi}_{t})}\big(\ln f({\mathbf{y}}^{\mathrm{dd},p},\pmb{\Psi})\big)\Big\},
\end{equation}
where $\mathbb{E}_{\sim q(\pmb{\Psi}_{t})}$ denotes expectation with respect to all latent variables except $\pmb{\Psi}_t$. Consequently, stable posterior approximations are obtained through alternating updates
\vspace{-2mm}
\begin{align}
q^{(i+1)}(\beta)\propto& \exp\Big(\mathbb{E}_{\sim q^{(i)}(\beta)}\big(\ln f({\mathbf{y}}^{\mathrm{dd},p},\pmb{\Psi})\big)\Big), \label{eq:q_beta_update_nsbl_otfs}\\
q^{(i+1)}({\mathbf{h}})\propto& \exp\Big(\mathbb{E}_{q^{(i+1)}(\beta)q^{(i)}(\pmb{\gamma})}\big(\ln f({\mathbf{y}}^{\mathrm{dd},p},\pmb{\Psi})\big)\Big), \label{eq:q_h_update_nsbl_otfs}\\
q^{(i+1)}(\pmb{\gamma})\propto& \exp\Big(\mathbb{E}_{q^{(i+1)}(\beta)q^{(i+1)}({\mathbf{h}})}\big(\ln f({\mathbf{y}}^{\mathrm{dd},p},\pmb{\Psi})\big)\Big). \label{eq:q_gamma_update_nsbl_otfs}
\end{align}
\vspace{-3mm}
The corresponding closed-form expressions are derived next.
\vspace{-5mm}
\subsection{Calculation of $q(\beta)$}
From Gamma prior on $\beta$ and \eqref{eq:likelihood_nsbl_otfs}, \eqref{eq:q_beta_update_nsbl_otfs} yields
\vspace{-2mm}
\begin{align}\label{eq:ln_q_beta_nsbl_otfs}
&\ln q^{(i+1)}(\beta)
\propto \ln f(\beta)+\mathbb{E}_{q^{(i)}({\mathbf{h}})}\big(\ln f({\mathbf{y}}^{\mathrm{dd},p}|{\mathbf{h}},\beta)\big)\nonumber\\
&\propto -\Big[d+\Big(\Vert {\mathbf{y}}^{\mathrm{dd},p}-{\boldsymbol{\Omega}}^{p}\pmb{\mu}_{h}^{(i)}\Vert_{2}^{2}
+\mathrm{Tr}\Big({\boldsymbol{\Omega}}^{p}\pmb{\Sigma}_{h}^{(i)}({\boldsymbol{\Omega}}^{p})^{H}\Big)\Big)\Big]\beta    \nonumber\\
&+\left(c+{MK_{2}N_{r}}-1\right)\ln\beta,
\end{align}
where $\pmb{\mu}_{h}^{(i)}=\mathbb{E}_{q^{(i)}({\mathbf{h}})}({\mathbf{h}})$ and
$\pmb{\Sigma}_{h}^{(i)}=\mathbb{E}_{q^{(i)}({\mathbf{h}})}\left(({\mathbf{h}}-\pmb{\mu}_{h}^{(i)})({\mathbf{h}}-\pmb{\mu}_{h}^{(i)})^{H}\right)$.
Thus, $q^{(i+1)}(\beta)$ is Gamma-distributed with parameters
\vspace{-2mm}
\begin{align}
&c^{(i+1)}=c+{MK_{2}N_{r}}, \nonumber \\
&d^{(i+1)}=d+\Vert {\mathbf{y}}^{\mathrm{dd},p}-{\boldsymbol{\Omega}}^{p}\pmb{\mu}_{h}^{(i)}\Vert_{2}^{2}
+\mathrm{Tr}\Big({\boldsymbol{\Omega}}^{p}\pmb{\Sigma}_{h}^{(i)}({\boldsymbol{\Omega}}^{p})^{H}\Big), \nonumber
\end{align}
namely, $q^{(i+1)}(\beta)=\mathrm{Gamma}\left(\beta|c^{(i+1)}, d^{(i+1)}\right).$
The posterior mean of $\beta$ is
\vspace{-3mm}
\begin{equation}\label{eq:beta_mean_nsbl_otfs}
\widehat{\beta}^{(i+1)}=\mathbb{E}_{q^{(i+1)}(\beta)}(\beta)=\frac{c^{(i+1)}}{d^{(i+1)}}.
\end{equation}
\vspace{-10mm}
\subsection{Calculation of $q({\mathbf{h}})$}
Using \eqref{eq:h_prior_nsbl_otfs} and \eqref{eq:q_h_update_nsbl_otfs}, we obtain
\vspace{-2mm}
\begin{align}\label{eq:ln_q_h_nsbl_otfs}
\ln ~ & q^{(i+1)}({\mathbf{h}})
\propto\ \mathbb{E}_{q^{(i+1)}(\beta)q^{(i)}(\pmb{\gamma})}\big(\ln f({\mathbf{y}}^{\mathrm{dd},p},\pmb{\Psi})\big)\nonumber\\
\propto&\ \mathbb{E}_{q^{(i+1)}(\beta)}\big(\ln f({\mathbf{y}}^{\mathrm{dd},p}|{\mathbf{h}},\beta)\big)
+\mathbb{E}_{q^{(i)}(\pmb{\gamma})}\big(\ln f({\mathbf{h}}|\pmb{\gamma})\big)\nonumber\\
\propto&\ -\widehat{\beta}^{(i+1)}||{\mathbf{y}}^{\mathrm{dd},p}-{\boldsymbol{\Omega}}^{p}{\mathbf{h}}||_{2}^{2}
-{\mathbf{h}}^{H}\mathbb{E}_{q^{(i)}(\pmb{\gamma})}(\pmb{\Gamma}){\mathbf{h}}.
\end{align}
Hence, $q^{(i+1)}({\mathbf{h}})=\mathcal{CN}\left({\mathbf{h}}|\pmb{\mu}_{h}^{(i+1)}, \pmb{\Sigma}_{h}^{(i+1)}\right),$
with
\vspace{-3mm}
\begin{align}
\pmb{\Sigma}_{h}^{(i+1)}&=
\left(\widehat{\beta}^{(i+1)}({\boldsymbol{\Omega}}^{p})^{H}{\boldsymbol{\Omega}}^{p}
+\mathbb{E}_{q^{(i)}(\pmb{\gamma})}(\pmb{\Gamma})\right)^{-1}, \label{eq:sigma_update_nsbl_otfs}\\
\pmb{\mu}_{h}^{(i+1)}&=
\widehat{\beta}^{(i+1)}\pmb{\Sigma}_{h}^{(i+1)}({\boldsymbol{\Omega}}^{p})^{H}{\mathbf{y}}^{\mathrm{dd},p}. \label{eq:mu_update_nsbl_otfs}
\end{align}
\vspace{-10mm}
\subsection{Calculation of $q(\pmb{\gamma})$}
Combining \eqref{eq:h_prior_nsbl_otfs} and \eqref{eq:gamma_prior_nsbl_otfs} in \eqref{eq:q_gamma_update_nsbl_otfs} gives
\vspace{-3mm}
\begin{align}\label{eq:ln_q_gamma_nsbl_otfs}
&\ln q^{(i+1)}(\pmb{\gamma})
\propto \ln f(\pmb{\gamma})+\mathbb{E}_{q^{(i+1)}({\mathbf{h}})}\big(\ln f({\mathbf{h}}|\pmb{\gamma})\big)\nonumber\\
&\propto \sum_{l=1}^{M_{\tau}G_{\nu}N_{t}N_{r}}
\left[a\ln(\gamma_l)-\gamma_l\Big(b+\mathbb{E}_{q^{(i+1)}({\mathbf{h}})}(|h_l|^{2})\Big)\right],
\end{align}
where $\mathbb{E}_{q^{(i+1)}({\mathbf{h}})}(|h_l|^{2})
=\left[\pmb{\Sigma}_{h}^{(i+1)}+\pmb{\mu}_{h}^{(i+1)}\big(\pmb{\mu}_{h}^{(i+1)}\big)^{H}\right]_{(l,l)}.$
Adopting the standard sparse setting $a=b=10^{-6}$ \cite{wipf2007empirical,liu2020uplink}, the parameter updates are
\vspace{-2mm}
\begin{equation}\label{eq:ab_update_nsbl_otfs}
a_{l}^{(i+1)}=a+1,\qquad b_{l}^{(i+1)}=b+\mathbb{E}_{q^{(i+1)}({\mathbf{h}})}(|h_l|^{2}),
\end{equation}
and therefore $q^{(i+1)}(\gamma_l)=\mathrm{Gamma}\left(\gamma_l|a_{l}^{(i+1)}, b_{l}^{(i+1)}\right),$
with mean $\widehat{\gamma}_{l}^{(i+1)}=\mathbb{E}_{q^{(i+1)}(\gamma_l)}(\gamma_l)=\frac{a_{l}^{(i+1)}}{b_{l}^{(i+1)}}.$
Hence, the expected precision matrix becomes
\vspace{-3mm}
\begin{align}\label{eq:EGamma_nsbl_otfs}
\mathbb{E}_{q^{(i+1)}(\pmb{\gamma})}(\pmb{\Gamma})
&=\mathrm{diag}\left(\mathbb{E}(\gamma_1),\ldots,\mathbb{E}\big(\gamma_{M_{\tau}G_{\nu}N_{t}N_{r}}\big)\right)\nonumber\\
&=\mathrm{diag}\left(\frac{a_{1}^{(i+1)}}{b_{1}^{(i+1)}},\ldots,
\frac{a_{M_{\tau}G_{\nu}N_{t}N_{r}}^{(i+1)}}{b_{M_{\tau}G_{\nu}N_{t}N_{r}}^{(i+1)}}\right).
\end{align}
By iterating the updates in \eqref{eq:ln_q_beta_nsbl_otfs}, \eqref{eq:ln_q_h_nsbl_otfs}, and \eqref{eq:ln_q_gamma_nsbl_otfs}, the variational posterior is refined, and the PA-VBI CIR estimate is obtained via the posterior mean
\vspace{-2mm}
\begin{equation}\label{eq:h_hat_nsbl_otfs}
\widehat{\mathbf{h}}_{\mathrm{PA-VBI}}=\mathbb{E}_{q^{(i+1)}({\mathbf{h}})}({\mathbf{h}})=\pmb{\mu}_{h}^{(i+1)}.
\end{equation}
The complete PA-VBI-based CE procedure for the AP-STS MIMO DCO-OTFS VLC system is given in Algorithm~\ref{vm_algo}.
\begin{algorithm}[t]
\small
\DontPrintSemicolon
\KwIn{Dictionary matrix ${\boldsymbol{\Omega}}^{p}\in \mathbb{C}^{MK_{2}N_{r}\times M_{\tau}G_{\nu}N_{t}N_{r}}$, received vector ${\mathbf{y}}^{\mathrm{dd},p}\in \mathbb{C}^{MK_{2}N_{r}\times 1}$, stopping parameters $m_{\max}$ and $\epsilon$}
\KwOut{Computed sparse DD-domain CIR vector $\widehat{\mathbf{h}}_{\mathrm{NSBL}}$}
\textbf{Initialization:} Parameters $a=b=c=d=10^{-6}$ and counter $i=0$

\While{$(\parallel\widehat{\boldsymbol{\gamma }}^{(i)} - \widehat{\boldsymbol{\gamma }}^{(i-1)}\parallel_2^2 > \epsilon~~ \&\&~~ i < m_{\max})$}
{
$i \leftarrow i+1$

Update the \textit{a posteriori} covariance and mean as
\begin{align*}
\pmb{\Sigma }_{h}^{(i+1)}&=\left(\widehat{\beta }^{(i+1)}({\boldsymbol{\Omega}}^{p})^{H}{\boldsymbol{\Omega}}^{p}
+\mathbb{E}_{q^{(i)}(\pmb{\gamma})}(\pmb{\Gamma})\right)^{-1},\\
\pmb{\mu }_{h}^{(i+1)}&=\widehat{\beta }^{(i+1)}\pmb{\Sigma }_{h}^{(i+1)}({\boldsymbol{\Omega}}^{p})^{H}{\mathbf{y}}^{\mathrm{dd},p}.
\end{align*}

Update the latent variables estimates as
\begin{align*}
\widehat{\beta }^{(i+1)}&=\mathbb{E}_{q^{(i+1)}(\beta)}(\beta)=\frac{c^{(i+1)}}{d^{(i+1)}},\\
{\pmb{\widehat\Gamma}}^{(i+1)}&=\mathbb{E}_{q^{(i+1)}(\pmb{\gamma})}(\pmb{\Gamma})
=\mathrm{diag}\left(\frac{a_{1}^{(i+1)}}{b_{1}^{(i+1)}},\ldots,
\frac{a_{M_{\tau}G_{\nu}N_{t}N_{r}}^{(i+1)}}{b_{M_{\tau}G_{\nu}N_{t}N_{r}}^{(i+1)}}\right).
\end{align*}
}\textbf{end}

\textbf{return:~~}{$\widehat{\mathbf{h}}_{\mathrm{PA-VBI}}=\mathbb{E}_{q^{(i+1)}(\mathbf{h})}(\mathbf{h})$}
\caption{PA-VBI-based sparse DD-domain CE for AP-STS MIMO DCO-OTFS VLC systems}
\label{vm_algo}
\end{algorithm}
\vspace{-3mm}
\section{Data aided joint CE and data detection for AP-STS MIMO DCO-OTFS VLC systems}
Starting from the decoupled DD-domain data output $\mathbf{Y}_{r}^{\mathrm{dd},d}$ in \eqref{eq:51}, separating the data contribution from the STS, substituting the DD-domain channel representation, and applying the $\mathrm{vec}(\cdot)$ operator yield
\vspace{-2mm}
\begin{align}
\mathbf{y}_{r}^{\mathrm{dd},d}
&=\sum_{t=1}^{N_t}\mathrm{vec}\left(\sum_{i=0}^{M_{\tau}-1}\sum_{j=0}^{G_{\nu}-1}
\mathbf{P}_{\mathrm{rx}}h_{i,j,r,t}(\bar{\boldsymbol{\Pi}})^{i}(\bar{\boldsymbol{\Delta}}_{i,j})\mathbf{P}_{\mathrm{tx}}\mathbf{S}_{t}^{d}\right) \nonumber \\
&+\mathbf{w}_{r}^{\mathrm{dd},d},
\label{eq:65}
\end{align}
where $\mathbf{y}_{r}^{\mathrm{dd},d}=\mathrm{vec}(\mathbf{Y}_{r}^{\mathrm{dd},d})\in\mathbb{C}^{MK_{1}\times 1}$ and
$\mathbf{w}_{r}^{\mathrm{dd},d}=\mathrm{vec}(\mathbf{W}_{r}^{\mathrm{dd},d})\in\mathbb{C}^{MK_{1}\times 1}$.
Equation \eqref{eq:65} can be given as
\vspace{-2mm}
\begin{equation}
\mathbf{y}_{r}^{\mathrm{dd},d}
=
\sum_{t=1}^{N_t}\sum_{i=0}^{M_{\tau}-1}\sum_{j=0}^{G_{\nu}-1}
\boldsymbol{\omega}_{i,j,t}^{d} h_{i,j,r,t}
+\mathbf{w}_{r}^{\mathrm{dd},d},
\label{eq:66}
\end{equation}
with $\boldsymbol{\omega}_{i,j,t}^{d}
=
\big(\mathbf{I}_{K_{1}}\otimes \mathbf{P}_{\mathrm{rx}}(\bar{\boldsymbol{\Pi}})^{i}(\bar{\boldsymbol{\Delta}}_{i,j})\mathbf{P}_{\mathrm{tx}}\big)\mathbf{s}_{t}^{d}$ and $\mathbf{s}_{t}^{d}=\mathrm{vec}(\mathbf{S}_{t}^{d}).$
Consequently, the observation model is given as
\vspace{-2mm}
\begin{equation}
\mathbf{y}_{r}^{\mathrm{dd},d}
=\sum_{t=1}^{N_t}\boldsymbol{\Omega}_{t}^{d}\mathbf{h}_{r,t}+\mathbf{w}_{r}^{\mathrm{dd},d},
\label{eq:67}
\end{equation}
where $\boldsymbol{\Omega}_{t}^{d}=\big[\boldsymbol{\omega}_{0,0,t}^{d} \cdots \boldsymbol{\omega}_{0,G_{\nu}-1,t}^{d} \cdots
\boldsymbol{\omega}_{M_{\tau}-1,0,t}^{d} \cdots$ $\boldsymbol{\omega}_{M_{\tau}-1,G_{\nu}-1,t}^{d}\big]\in\mathbb{C}^{MK_{1}\times M_{\tau}G_{\nu}}$ is the dictionary associated with the $t$th LED.
By aggregating the LED-specific dictionaries, \eqref{eq:67} becomes
\vspace{-3mm}
\begin{equation}
\mathbf{y}_{r}^{\mathrm{dd},d}
=
\widetilde{\boldsymbol{\Omega}}^{d}\mathbf{h}_{r}
+\mathbf{w}_{r}^{\mathrm{dd},d},
\label{eq:68}
\end{equation}
where $\widetilde{\boldsymbol{\Omega}}^{d}=[\boldsymbol{\Omega}_{1}^{d},\boldsymbol{\Omega}_{2}^{d},\ldots,\boldsymbol{\Omega}_{N_{t}}^{d}]
\in\mathbb{C}^{MK_{1}\times M_{\tau}G_{\nu}N_{t}}$ and
$\mathbf{h}_{r}=[\mathbf{h}_{r,1}^{T},\mathbf{h}_{r,2}^{T},\ldots,\mathbf{h}_{r,N_{t}}^{T}]^{T}\in\mathbb{R}_{+}^{M_{\tau}G_{\nu}N_{t}\times 1}$.
Stacking $\mathbf{y}_{r}^{\mathrm{dd},d}$ over all PDs yields $\mathbf{y}^{\mathrm{dd},d}
=[(\mathbf{y}_{1}^{\mathrm{dd},d})^{T},(\mathbf{y}_{2}^{\mathrm{dd},d})^{T},\ldots,(\mathbf{y}_{N_{r}}^{\mathrm{dd},d})^{T}]^{T}
\in\mathbb{C}^{MK_{1}N_{r}\times 1},$
which can be expressed as
\vspace{-2mm}
\begin{equation}
\mathbf{y}^{\mathrm{dd},d}
=\boldsymbol{\Omega}^{d}\mathbf{h}+\mathbf{w}^{d},
\label{eq:69}
\end{equation}
where $\mathbf{h}=[\mathbf{h}_{1}^{T},\mathbf{h}_{2}^{T},\ldots,\mathbf{h}_{N_{r}}^{T}]^{T}\in \mathbb{R}_{+}^{M_{\tau}G_{\nu}N_{t}N_r \times 1},$ $\boldsymbol{\Omega}^{d}
=\mathbf{I}_{N_r}\otimes \widetilde{\boldsymbol{\Omega}}^{d}\in\mathbb{C}^{MK_{1}N_r\times M_{\tau}G_{\nu}N_tN_r},$ $\mathbf{w}^{d}
=[(\mathbf{w}_{1}^{\mathrm{dd},d})^{T},(\mathbf{w}_{2}^{\mathrm{dd},d})^{T},\ldots,(\mathbf{w}_{N_{r}}^{\mathrm{dd},d})^{T}]^{T}\in\mathbb{C}^{MK_{1}N_{r}\times 1}.$
For data-aided AP-STS-based MIMO DCO-OTFS CSI estimation, the pilot-only model in \eqref{eq:62} and the data model in \eqref{eq:69} can be combined to form the joint observation
\vspace{-2mm}
\begin{align}
\underbrace{\begin{bmatrix}
{\mathbf{y}}^{\mathrm{dd},d}\\
{\mathbf{y}}^{\mathrm{dd},p}
\end{bmatrix}}_{{\mathbf{y}}\in\mathbb{C}^{MN_aN_r\times 1}}
=
\underbrace{\begin{bmatrix}
\boldsymbol{\Omega}^{d}\\
\boldsymbol{\Omega}^{p}
\end{bmatrix}}_{\boldsymbol{\Phi}\in\mathbb{C}^{MN_aN_r\times M_{\tau}G_{\nu}N_{t}N_r}}
\mathbf{h}
+
\underbrace{\begin{bmatrix}
\mathbf{w}^{d}\\
\mathbf{w}^{p}
\end{bmatrix}}_{\mathbf{v}\in\mathbb{C}^{MN_aN_r\times 1}}.
\end{align}
Accordingly, the compact data-aided model is
\vspace{-3mm}
\begin{equation}\label{po1}
\mathbf{y}=\boldsymbol{\Phi}\mathbf{h}+\mathbf{v},
\end{equation}
where the noise covariance is $\mathbf{R}_{v}=\mathrm{blkdiag}(\mathbf{R}_{w_{d}},\mathbf{R}_{w_{p}})\in\mathbb{C}^{MN_aN_r\times MN_aN_r}$. The ensuing development details the proposed DA-VBI procedure for joint CE and data detection.

This section presents a DA-VBI framework for the AP-STS MIMO DCO-OTFS VLC system, with the objective of enhancing DD-domain CSI estimation by exploiting the unknown data symbols as virtual pilots. In contrast to pilot-only methods, the data-bearing DD-domain symbols are treated as latent variables and are inferred jointly with the channel and its hyperparameters. As a result, the channel estimate is updated using both the pilot observations and the soft information extracted from the data component, while symbol detection is performed through probabilistic inference. Although the stacked model in \eqref{po1} increases the number of measurements, the data-dependent part of $\boldsymbol{\Phi}$ cannot be constructed \textit{a priori}. To address this coupling, we introduce the DD-domain data vector $\mathbf{s}^{d}=\mathrm{vec}(\widetilde{\mathbf{S}}^{d})\in \mathbb{C}^{MK_1N_t\times 1}$ as an additional hidden variable and estimate it in a probabilistic manner. Hence, the latent-variable set for DA-VBI is defined as $\pmb{\Psi}'=\{\mathbf{s}^{d},\beta,\mathbf{h},\pmb{\gamma}\}$. The key mechanism is that the posterior of $\mathbf{s}^{d}$ provides soft symbol estimates, which are then used to update the sensing matrix and refine the CSI estimate. Let the transmitted symbols be drawn from a $P$-QAM constellation $\mathcal{A}_{P}=\{a_{1},\ldots,a_{P}\}$. With equiprobable signaling and independent symbols, each entry satisfies $p(s^{d}_{n})=\frac{1}{P}$ for $s^{d}_{n}\in\mathcal{A}_{P}$, and the prior factorizes as $p(\mathbf{s}^{d})=\prod_{n} p(s^{d}_{n})$, where the product spans all DD-domain data entries that contribute to $\boldsymbol{\Omega}^{d}$. Under a mean-field approximation, the variational posterior is factorized as
\vspace{-2mm}
\begin{equation}
q(\pmb{\Psi}')=q(\mathbf{s}^{d})q(\beta)q(\mathbf{h})q(\pmb{\gamma}).
\end{equation}
The corresponding alternating updates are
\vspace{-2mm}
\begin{align}
q^{(i+1)}(\mathbf{s}^{d})
&\propto
\exp\Big(
\mathbb{E}_{q^{(i)}(\beta)q^{(i)}(\mathbf{h})q^{(i)}(\pmb{\gamma})}\big( \ln p(\mathbf{y},\pmb{\Psi}')\big)
\Big), \label{eq:q_s_update} \\
q^{(i+1)}(\beta)
&\propto
\exp \Big(
\mathbb{E}_{q^{(i+1)}(\mathbf{s}^{d})q^{(i)}(\mathbf{h})q^{(i)}(\pmb{\gamma})}\big( \ln p(\mathbf{y},\pmb{\Psi}')\big)
\Big), \label{eq:q_beta_update_da} \\
q^{(i+1)}(\mathbf{h})
&\propto
\exp \Big(
\mathbb{E}_{q^{(i+1)}(\mathbf{s}^{d})q^{(i+1)}(\beta)q^{(i)}(\pmb{\gamma})}\big( \ln p(\mathbf{y},\pmb{\Psi}')\big)
\Big), \label{eq:q_h_update_da} \\
q^{(i+1)}(\pmb{\gamma})
&\propto
\exp \Big(
\mathbb{E}_{q^{(i+1)}(\mathbf{s}^{d})q^{(i+1)}(\beta)q^{(i+1)}(\mathbf{h})}\big( \ln p(\mathbf{y},\pmb{\Psi}')\big)
\Big). \label{eq:q_gamma_update_da}
\end{align}
These iterations yield soft estimates of $\mathbf{s}^{d}$, which are subsequently used to update the data-dependent sensing matrix and to iteratively refine the sparse DD-domain CSI estimate.
\vspace{-3mm}

\subsection{Update of $q(\mathbf{s}^{d})$}
Using the vectorized representation of \eqref{eq:53}, we have 
\vspace{-2mm}
\begin{equation}
\mathbf{y}^{\mathrm{dd},d}=\widetilde{\mathbf{H}}\mathbf{s}^{d}+\mathbf{w}^d,
\end{equation}
where $\widetilde{\mathbf{H}} = \mathbf{I}_{K_1} \otimes \widetilde{\mathbf{H}}^{\mathrm{dd}} $. Thus, 
the update of $q(\mathbf{s}^{d})$ follows as
\vspace{-3mm}
\begin{align}
q^{(i+1)}(\mathbf{s}^{d})
&\propto
\exp\Big(
\mathbb{E}_{q^{(i)}(\beta)q^{(i)}(\mathbf{h})}\big(
-\beta \Vert \mathbf{y}^{\mathrm{dd},d}-\widetilde{\mathbf{H}}\mathbf{s}^{d}\Vert_{2}^{2}
\big)
\Big)p(\mathbf{s}^{d}) \nonumber\\
&\propto
\exp\left(
-(\mathbf{s}^{d}-\mathbf{u}_{\mathbf{s}})^{H}\boldsymbol{\Sigma }_{\mathbf{s}}^{-1}(\mathbf{s}^{d}-\mathbf{u}_{\mathbf{s}})
\right)p(\mathbf{s}^{d}),
\end{align}
where $\mathbf{u}_{\mathbf{s}}
=\widehat{\beta}^{(i)} \boldsymbol{\Sigma }_{\mathbf{s}}
\Big(\mathbb{E}_{q^{(i)}(\mathbf{h})}(\widetilde{\mathbf{H}})\Big)^{H}\mathbf{y}^{\mathrm{dd},d}$, and $\boldsymbol{\Sigma }_{\mathbf{s}}
=\Big(\widehat{\beta}^{(i)}
\mathbb{E}_{q^{(i)}(\mathbf{h})}\big(\widetilde{\mathbf{H}}^{H}\widetilde{\mathbf{H}}\big)\Big)^{-1}.$
The required terms $\mathbb{E}_{q^{(i)}(\mathbf{h})}(\widetilde{\mathbf{H}})$ and $\mathbb{E}_{q^{(i)}(\mathbf{h})}(\widetilde{\mathbf{H}}^{H}\widetilde{\mathbf{H}})$ are provided in Appendix B. Denoting by $\mathcal{S}_{j}$ the $j$th candidate vector formed from $P$-QAM combinations across $Q$ entries, the discrete posterior is given as
\vspace{-4mm}
\begin{equation}
q^{(i+1)}(\mathbf{s}^{d})
=
\begin{cases}
m n_{1} p_{1}', & \mathbf{s}^{d}=\mathcal{S}_{1} \\
m n_{2} p_{2}', & \mathbf{s}^{d}=\mathcal{S}_{2} \\
\vdots \\
m n_{P^{Q}} p_{P^{Q}}', & \mathbf{s}^{d}=\mathcal{S}_{P^{Q}},
\end{cases}
\end{equation}
with $n_{j}=\exp\left(
-(\mathcal{S}_{j}-\mathbf{u}_{\mathbf{s}})^{H}\boldsymbol{\Sigma }_{\mathbf{s}}^{-1}(\mathcal{S}_{j}-\mathbf{u}_{\mathbf{s}})
\right)$, $m=\Big(\sum_{j=1}^{P^{Q}} n_{j} p_{j}'\Big)^{-1}.$
Thus, posterior mean and covariance are
\vspace{-3mm}
\begin{align}\label{fg1}
\mathbf{u}_{\mathbf{s}}'=\sum_{j=1}^{P^{Q}} m n_{j} p_{j}'\mathcal{S}_{j}, ~
\boldsymbol{\Sigma }_{\mathbf{s}}'=\sum_{j=1}^{P^{Q}}
(\mathcal{S}_{j}-\mathbf{u}_{\mathbf{s}}')(\mathcal{S}_{j}-\mathbf{u}_{\mathbf{s}}')^{H} m n_{j} p_{j}'.
\end{align}
\vspace{-10mm}

\subsection{Update of $q(\beta)$}
Using \eqref{po1} together with the Gamma prior on $\beta$, the update can be expressed as
\vspace{-3mm}
\begin{align}
&\ln q^{(i+1)}(\beta)
\propto
\mathbb{E}_{q^{(i+1)}(\mathbf{s}^{d})q^{(i)}(\mathbf{h})q^{(i)}(\pmb{\gamma})}
\Big(\ln p(\mathbf{y},\pmb{\Psi}')\Big)\nonumber\\
&\propto
\mathbb{E}_{q^{(i+1)}(\mathbf{s}^{d})q^{(i)}(\mathbf{h})}
\Big(\ln p(\mathbf{y}| \mathbf{s}^{d},\beta,\mathbf{h})\Big)
+\ln p(\beta)\nonumber\\
&\propto
(c+MN_aN_r-1)\ln(\beta)-\beta\Big[\mathrm{tr}\Big\{
\mathbb{E}_{q^{(i+1)}(\mathbf{s}^{d})}\Big(\boldsymbol{\Phi}\pmb{\Sigma}_{\mathbf{h}}^{(i)}\boldsymbol{\Phi}^{H}\Big)\Big\} \nonumber\\
&+d+\mathbf{y}^{H}\mathbf{y}
-2\mathrm{Re}\Big(\mathbf{y}^{H}\mathbb{E}_{q^{(i+1)}(\mathbf{s}^{d})}\big(\boldsymbol{\Phi}\big)\pmb{\mu}_{\mathbf{h}}^{(i)}\Big) +\big(\pmb{\mu}_{\mathbf{h}}^{(i)}\big)^{H} \nonumber\\
&\times\mathbb{E}_{q^{(i+1)}(\mathbf{s}^{d})}\big(\boldsymbol{\Phi}^{H}\boldsymbol{\Phi}\big)
\pmb{\mu}_{\mathbf{h}}^{(i)}
\Big]. 
\end{align}
Compared with the pilot-only update, the matrices $\mathbb{E}_{q^{(i+1)}(\mathbf{s}^{d})}(\boldsymbol{\Phi})$,
$\mathbb{E}_{q^{(i+1)}(\mathbf{s}^{d})}(\boldsymbol{\Phi}^{H}\boldsymbol{\Phi})$, and
$\mathrm{Tr}\Big(\mathbb{E}_{q^{(i+1)}(\mathbf{s}^{d})}(\boldsymbol{\Phi}\pmb{\Sigma}_{\mathbf{h}}^{(i)}\boldsymbol{\Phi}^{H})\Big)$
replace their pilot-only counterparts, and their evaluation is summarized in Appendix C. Hence, $\beta$ follows a Gamma distribution with $c_{\beta}^{(i+1)}=c+MN_aN_r$, $d_{\beta}^{(i+1)}=d+\mathbf{y}^{H}\mathbf{y} -2\mathrm{Re}\Big(\mathbf{y}^{H}\mathbb{E}_{q^{(i+1)}(\mathbf{s}^{d})}\big(\boldsymbol{\Phi}\big)\pmb{\mu}_{\mathbf{h}}^{(i)}\Big)
+\mathrm{tr}\Big\{
\mathbb{E}_{q^{(i+1)}(\mathbf{s}^{d})}\Big(\boldsymbol{\Phi}\pmb{\Sigma}_{\mathbf{h}}^{(i)}\boldsymbol{\Phi}^{H}\Big)\Big\}$ $
+\big(\pmb{\mu}_{\mathbf{h}}^{(i)}\big)^{H}
\mathbb{E}_{q^{(i+1)}(\mathbf{s}^{d})}\big(\boldsymbol{\Phi}^{H}\boldsymbol{\Phi}\big)
\pmb{\mu}_{\mathbf{h}}^{(i)},$
and $q^{(i+1)}(\beta)=\mathrm{Gamma} \left(\beta|c_{\beta}^{(i+1)},d_{\beta}^{(i+1)}\right). $
Thus, the mean of $\beta$ is given as
\vspace{-3mm}
\begin{equation}
\widehat{\beta}^{(i+1)}=\mathbb{E}_{q^{(i+1)}(\beta)}(\beta)=\frac{c_{\beta}^{(i+1)}}{d_{\beta}^{(i+1)}}. 
\end{equation}
\vspace{-8mm}

\subsection{Update of $q(\mathbf{h})$}
Based on \eqref{po1}, we arrive at
\vspace{-2mm}
\begin{align}
&\ln q^{(i+1)}(\mathbf{h})
\propto
\mathbb{E}_{q^{(i+1)}(\mathbf{s}^{d})q^{(i+1)}(\beta)q^{(i)}(\pmb{\gamma})}
\big(\ln p(\mathbf{y},\pmb{\Psi}')\big)\nonumber\\
\propto&
\mathbb{E}_{q^{(i+1)}(\mathbf{s}^{d})q^{(i+1)}(\beta)}
\big(\ln p(\mathbf{y}|\mathbf{s}^{d},\beta,\mathbf{h})\big)
+\mathbb{E}_{q^{(i)}(\pmb{\gamma})}
\big(\ln p(\mathbf{h}|\pmb{\gamma})\big)\nonumber\\
\propto&-\widehat{\beta}^{(i+1)}
\big(\mathbf{h}^{H}\mathbb{E}_{q^{(i+1)}(\mathbf{s}^{d})}\big(\boldsymbol{\Phi}^{H}\boldsymbol{\Phi}\big)\mathbf{h}+\mathbf{y}^{H}\mathbf{y} \nonumber\\
&-2\mathrm{Re}\big(\mathbf{y}^{H}\mathbb{E}_{q^{(i+1)}(\mathbf{s}^{d})}(\boldsymbol{\Phi})\mathbf{h}\big)
\big)-\mathbf{h}^{H}\mathbb{E}_{q^{(i)}(\pmb{\gamma})}\big(\pmb{\Gamma}\big)\mathbf{h}.
\end{align}
Therefore, $\mathbf{h}$ follows the Gaussian distribution, yielding
\vspace{-2mm}
\begin{equation}
q^{(i+1)}(\mathbf{h})=\mathcal{CN}\left(\mathbf{h}|\pmb{\mu}_{\mathbf{h}}^{(i+1)},\pmb{\Sigma}_{\mathbf{h}}^{(i+1)}\right),
\end{equation}
where $\pmb{\Sigma}_{\mathbf{h}}^{(i+1)}
=\left(\widehat{\beta}^{(i+1)}\mathbb{E}_{q^{(i+1)}(\mathbf{s}^{d})}\big(\boldsymbol{\Phi}^{H}\boldsymbol{\Phi}\big)+\mathbb{E}_{q^{(i)}(\pmb{\gamma})}\big(\pmb{\Gamma}\big)
\right)^{-1},$ and $\pmb{\mu}_{\mathbf{h}}^{(i+1)}
=\widehat{\beta}^{(i+1)}
\pmb{\Sigma}_{\mathbf{h}}^{(i+1)}
\mathbb{E}_{q^{(i+1)}(\mathbf{s}^{d})}\big(\boldsymbol{\Phi}^{H}\big)
\mathbf{y}.$
Here, $\pmb{\Gamma}=\mathrm{diag}(\gamma_{1},\gamma_{2},\ldots,\gamma_{M_{\tau}G_{\nu}N_{t}N_{r}})$, and $\widehat{\beta}^{(i+1)}=\mathbb{E}_{q^{(i+1)}(\beta)}(\beta)$.
\vspace{-6mm}
\subsection{Update of $q(\pmb{\gamma})$}
The update of $q(\pmb{\gamma})$ follows the same steps as in the pilot-aided case and is therefore omitted for brevity.

Algorithm~\ref{algo_da_nsbl_emstyle} summarizes the resulting joint CE and data detection procedure. The initialization uses the pilot-aided expressions for $\widehat{\beta}^{(0)}$, $\pmb{\mu}_{\mathbf{h}}^{(0)}$, $\pmb{\Sigma}_{\mathbf{h}}^{(0)}$, and $\pmb{\gamma}^{(0)}$. The known portion of the initial sensing matrix $\boldsymbol{\Phi}$ is constructed from pilot symbols, while the unknown data-dependent entries are initialized to zero. Since $\mathbf{u}'_{\mathbf{s}}$ generally requires multiple iterations to stabilize, inner iterations are executed by updating the remaining latent variables using $\boldsymbol{\Phi}$ built from $\mathbf{u}_{\mathbf{s}}^{\prime (i)}$ and $\boldsymbol{\Sigma}_{\mathbf{s}}^{\prime (i)}$ from the previous outer iteration. The inner loop terminates when $\frac{\Vert \pmb{\mu}_{\mathbf{s}}^{\prime (i+1)}-\pmb{\mu}_{\mathbf{s}}^{\prime (i)}\Vert_{2}^{2}}
{\Vert \pmb{\mu}_{\mathbf{s}}^{\prime (i)}\Vert_{2}^{2}}$
falls below a prescribed threshold or when the maximum allowable number of iterations is reached. The outer loop subsequently updates $\widehat{\beta}$, $\pmb{\mu}_{\mathbf{h}}$, $\pmb{\Sigma}_{\mathbf{h}}$, and $\mathbb{E}_{q(\pmb{\gamma})}(\pmb{\Gamma})$, and terminates when $\frac{\Vert\pmb{\gamma}^{(i+1)}-\pmb{\gamma}^{(i)}\Vert_{2}^{2}}
{\Vert\pmb{\gamma}^{(i)}\Vert_{2}^{2}}$
satisfies the stopping criterion.
\vspace{-5mm}

\subsection{BCRLB for MIMO DCO-OTFS VLC Systems}
The BCRLB is adopted to characterize a fundamental lower bound on the mean-square-error (MSE) attainable when estimating the sparse CSI vector $\widehat{\mathbf{h}}$. To obtain this bound, we form the Bayesian Fisher information matrix (BFIM) $\mathbf{T}_B\in\mathbb{C}^{M_{\tau}G_{\nu}N_tN_r \times M_{\tau}G_{\nu}N_tN_r}$ associated with the concatenated CIR vector $\mathbf{h}$. The BFIM decomposes as $\mathbf{T}_B=\mathbf{T}_P+\mathbf{T}_D,$
where $\mathbf{T}_P\in\mathbb{C}^{M_{\tau}G_{\nu}N_tN_r\times M_{\tau}G_{\nu}N_tN_r}$ and $\mathbf{T}_D\in\mathbb{C}^{M_{\tau}G_{\nu}N_tN_r\times M_{\tau}G_{\nu}N_tN_r}$ denote the information contributions due to the prior distribution of $\mathbf{h}$ and the pilot-observation model for $\mathbf{y}$, respectively. These matrices are defined by $\mathbf{T}_{P}=-\mathbb{E}_{\mathbf{h},\pmb{\gamma}}
\left\{
\frac{\partial^{2}\mathcal{L}(\mathbf{h};\pmb{\gamma})}{\partial \mathbf{h}\partial \mathbf{h}^{H}}
\right\}$, and $\mathbf{T}_{D}=-\mathbb{E}_{\mathbf{y},\mathbf{h},\beta}
\left\{
\frac{\partial^{2}\mathcal{L}(\mathbf{y}\vert \mathbf{h},\beta)}{\partial \mathbf{h}\partial \mathbf{h}^{H}}
\right\},$
where $\mathcal{L}(\mathbf{h};\pmb{\gamma})=\log f(\mathbf{h};\pmb{\gamma})$ is the log-prior of $\mathbf{h}$ and $\mathcal{L}(\mathbf{y}\vert \mathbf{h},\beta)=\log f(\mathbf{y}\vert \mathbf{h},\beta)$ is the log-likelihood for the received observations. These quantities can be written as
\vspace{-2mm}
\begin{align}
\mathcal{L}(\mathbf{y}\vert \mathbf{h},\beta)
&=\kappa_{1}-\beta(\mathbf{y}-\mathbf{\Phi}\mathbf{h})^{H}(\mathbf{y}-\mathbf{\Phi}\mathbf{h}),\\
\mathcal{L}(\mathbf{h};\pmb{\gamma})
&=\kappa_{2}-\mathbf{h}^{H}\mathbf{\Gamma}\mathbf{h},
\end{align}
where $\kappa_{1}=MN_aN_r\log(\beta)-MN_aN_r\log(\pi)$ and $\kappa_{2}=\log(\det(\mathbf{\Gamma}))-M_{\tau}G_{\nu}N_tN_r\log(\pi)$ are constants with respect to $\mathbf{h}$. Simplifying the above expressions yields the closed-form results $\mathbf{T}_D=\mathbb{E}_{q^{(i+1)}(\beta)}(\beta)\mathbf{\Phi}^{H}\mathbf{\Phi},$ and $\mathbf{T}_P=\mathbb{E}_{q^{(i+1)}(\pmb{\gamma})}(\mathbf{\Gamma}).$
Consequently, the BFIM is given by
\vspace{-2mm}
\begin{equation}
\mathbf{T}_B=
\mathbb{E}_{q^{(i+1)}(\pmb{\gamma})}(\mathbf{\Gamma})+
\mathbb{E}_{q^{(i+1)}(\beta)}(\beta)\mathbf{\Phi}^{H}\mathbf{\Phi}.
\end{equation}
Therefore, for the MSE defined as $\mathrm{MSE}=\mathbb{E}\big[\Vert \mathbf{h}-\widehat{\mathbf{h}}\Vert_{2}^{2}\big]$, the BCRLB satisfies
\vspace{-2mm}
\begin{align}\label{BCRLB1}
&\mathrm{MSE}(\widehat{\mathbf{h}})
\geq \mathrm{Tr}\big(\mathbf{T}_B^{-1}\big) \nonumber\\
&=\mathrm{Tr}\left(
\left(
\mathbb{E}_{q^{(i+1)}(\pmb{\gamma})}(\mathbf{\Gamma})
+
\mathbb{E}_{q^{(i+1)}(\beta)}(\beta)\,\mathbf{\Phi}^{H}\mathbf{\Phi}
\right)^{-1}
\right).
\end{align}

\begin{algorithm}[t]
\small
\DontPrintSemicolon
\KwIn{Stacked observation vector $\mathbf{y}$, pilot dictionary $\boldsymbol{\Omega}^{p}$, constellation $\mathcal{A}_{P}$, stopping parameters $\epsilon$ and $m_{\max}$}
\KwOut{Estimated CSI vector $\widehat{\mathbf{h}}_{\mathrm{DA-VBI}}=\pmb{\mu}_{\mathbf{h}}^{(m)}$ and detected DD-domain data $\widehat{\mathbf{s}}^{d,(m)}$}
\textbf{Initialization:} $a=b=c=d=10^{-6}$, $m=0$. Initialize $\widehat{\beta}^{(0)}$, $\pmb{\mu}_{\mathbf{h}}^{(0)}$, $\pmb{\Sigma}_{\mathbf{h}}^{(0)}$, and $\widehat{\pmb{\gamma}}^{(0)}$ from the PA-VBI solution. Set $\widehat{\mathbf{s}}^{d,(0)}=\mathbf{0}$ on data-bearing entries and enforce pilots. Construct $(\boldsymbol{\Omega}^{d})^{(0)}$ and
$\boldsymbol{\Phi}^{(0)}=\begin{bmatrix}
(\boldsymbol{\Omega}^{d})^{(0)}\\[0.5mm]
\boldsymbol{\Omega}^{p}
\end{bmatrix}$.

\While{$\left(||\widehat{\pmb{\gamma}}^{(m)}-\widehat{\pmb{\gamma}}^{(m-1)}||_{2}>\epsilon~~\&\&~~ m<m_{\max}\right)$}
{
$m\leftarrow m+1$

\textbf{E-step:} Update the \textit{a posteriori} covariance and mean as\\
$\pmb{\Sigma}_{\mathbf{h}}^{(m)}=\Big(\widehat{\beta}^{(m-1)}\mathbb{E}_{q^{(m-1)}(\mathbf{s}^{d})}\big((\boldsymbol{\Phi}^{(m-1)})^{H}\boldsymbol{\Phi}^{(m-1)}\big)$\\ \hspace{8mm}$+\mathbb{E}_{q^{(m-1)}(\pmb{\gamma})}(\pmb{\Gamma})\Big)^{-1}$,\\
$\pmb{\mu}_{\mathbf{h}}^{(m)}=\widehat{\beta}^{(m-1)}\pmb{\Sigma}_{\mathbf{h}}^{(m)}\mathbb{E}_{q^{(m-1)}(\mathbf{s}^{d})}\big((\boldsymbol{\Phi}^{(m-1)})^{H}\big)\mathbf{y}$.

\textbf{M-step 1:} Update the latent variables as\\
$\widehat{\beta}^{(m)}=\mathbb{E}_{q^{(m)}(\beta)}(\beta)=\dfrac{c_{\beta}^{(m)}}{d_{\beta}^{(m)}}$,\\
${\pmb{\widehat\Gamma}}^{(m)}
=\mathrm{diag}\left(\widehat{\gamma}_{1}^{(m)},\ldots,\widehat{\gamma}_{M_{\tau}G_{\nu}N_{t}N_{r}}^{(m)}\right)$.

\textbf{M-step 2:} \\
$1)$ Update $q(\mathbf{s}^{d})$ and obtain $\pmb{\mu}_{\mathbf{s}}^{\prime (m)}$, $\pmb{\Sigma}_{\mathbf{s}}^{\prime (m)}$ using \eqref{fg1}.\\ 
$2)$ Demodulate $\pmb{\mu}_{\mathbf{s}}^{\prime (m)}$ to $\mathcal{A}_{P}$ and get $\widehat{\mathbf{s}}^{d,(m)}$.\\
$3)$ Construct $(\boldsymbol{\Omega}^{d})^{(m)}$ and update $\boldsymbol{\Phi}^{(m)}$ using \eqref{eq:69}-\eqref{po1}.

}\textbf{end}

\textbf{return:~~}{$\widehat{\mathbf{h}}_{\mathrm{DA-VBI}}=\mathbb{E}_{q^{(m)}(\mathbf{h})}(\mathbf{h})=\pmb{\mu}_{\mathbf{h}}^{(m)}$ and $\widehat{\mathbf{s}}^{d,(m)}$}
\caption{DA-VBI-based joint CE and data detection for AP-STS MIMO DCO-OTFS VLC systems}
\label{algo_da_nsbl_emstyle}
\end{algorithm}

\begin{figure*}[ht]
\centering
\captionsetup[subfigure]{justification=centering}
\subfloat[]{\label{1}\includegraphics[width=60mm,height=50mm]{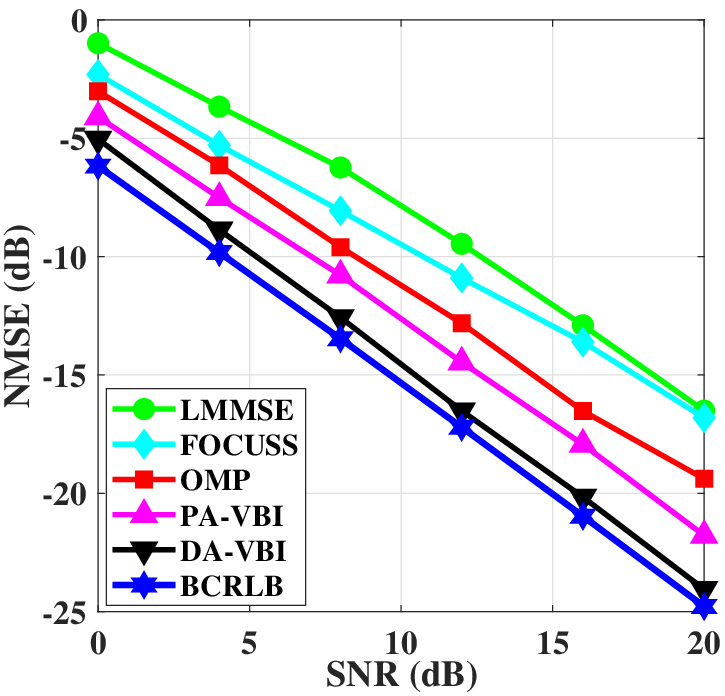}} 
\subfloat[]{\label{2}\includegraphics[width=60mm,height=50mm]{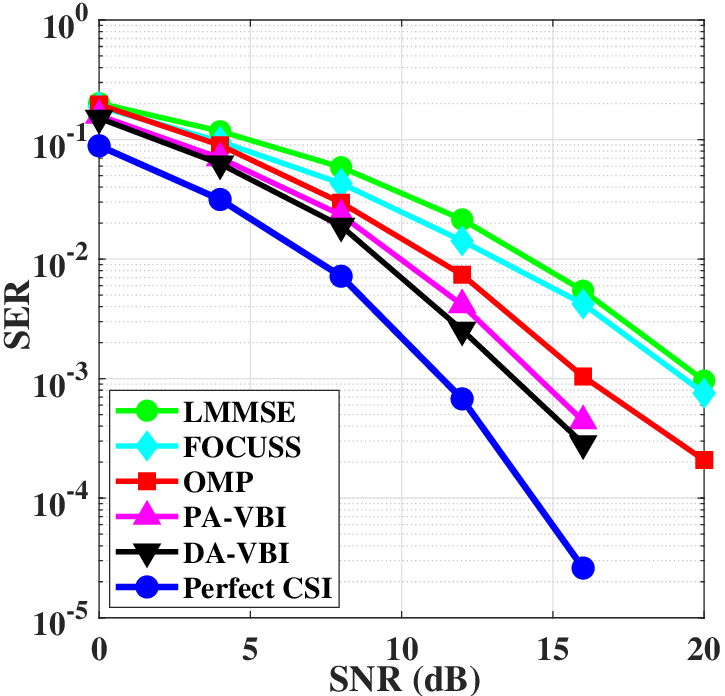}}
\subfloat[]{\label{3}\includegraphics[width=60mm,height=50mm]{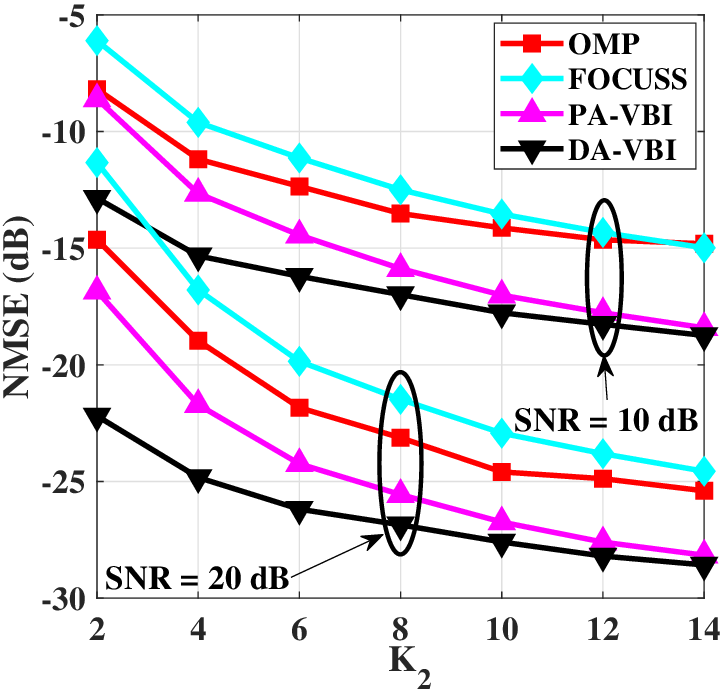}}
\caption{AP-STS MIMO DCO-OTFS VLC system, demonstrates (a) NMSE versus SNR performance;  (b) SER versus SNR performance; (c) NMSE versus pilot length ($K_2$) performance at SNR $= 10$ and $20$ dB.}
\label{a}
\end{figure*}

\vspace{-4mm}
\section{Simulation Results}
This section provides a comparative performance evaluation of the proposed PA-VBI and DA-VBI schemes against existing CIR estimation methods, namely FOCUSS, OMP, and LMMSE \cite{saxena2023sparse}, for CSI recovery in AP-STS MIMO DCO-OTFS VLC systems. The comparison is carried out in terms of SER, NMSE, and pilot length. In particular, SER reflects the detection reliability at the receiver based on the estimated CIR. In the simulation setup, $\mathbf{R}_h=\mathbf{I}_{M_\tau N_\nu N_t N_r}$, and the SNR in dB is defined as $\mathrm{SNR\ (dB)}=10\log_{10}(\beta)$. The PA-VBI and DA-VBI algorithms employ the convergence parameters $\epsilon=10^{-6}$ and $m_{\max}=50$, whereas the OMP threshold is set to $\xi=0.1$. For FOCUSS, the configuration uses the noise regularization factor $\sigma^2$, the $l_p$-norm with $p=0.8$, a stopping tolerance of $10^{-5}$, and a maximum of $800$ iterations. The remaining simulation parameters are $\Delta f=240$ kHz, $M=64$, $N=64$, $K_2=4$, $L=16$, $L_p=5$, $M_\tau=16$, $N_\nu=15$, $N_r=2$, $N_t=2$, $B_{\text{DC}}=7$ dB, BPSK modulation, and a rectangular pulse shape \cite{saxena2023sparse,liao2023sparse,xu2023optical}.

\begin{figure*}[ht]
\centering
\captionsetup[subfigure]{justification=centering}
\subfloat[]{\label{4}\includegraphics[width=60mm,height=50mm]{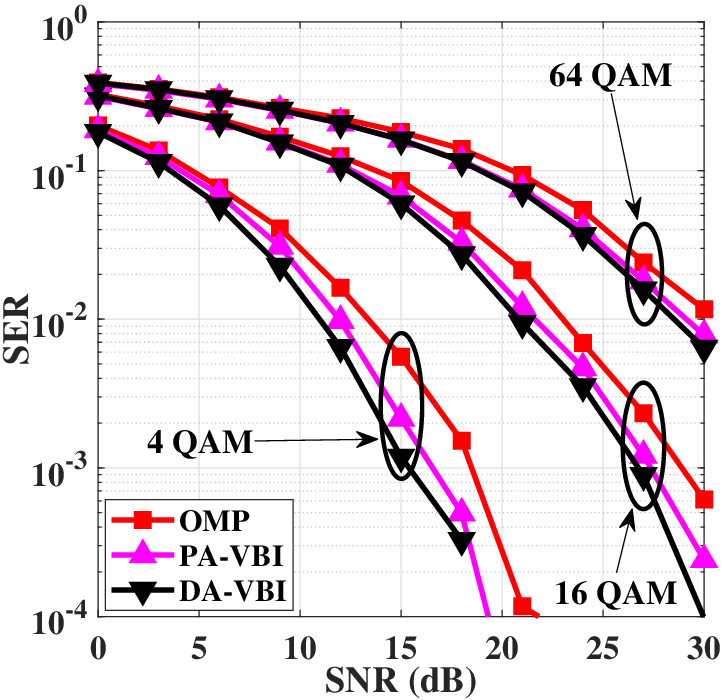}}
\subfloat[]{\label{5}\includegraphics[width=60mm,height=50mm]{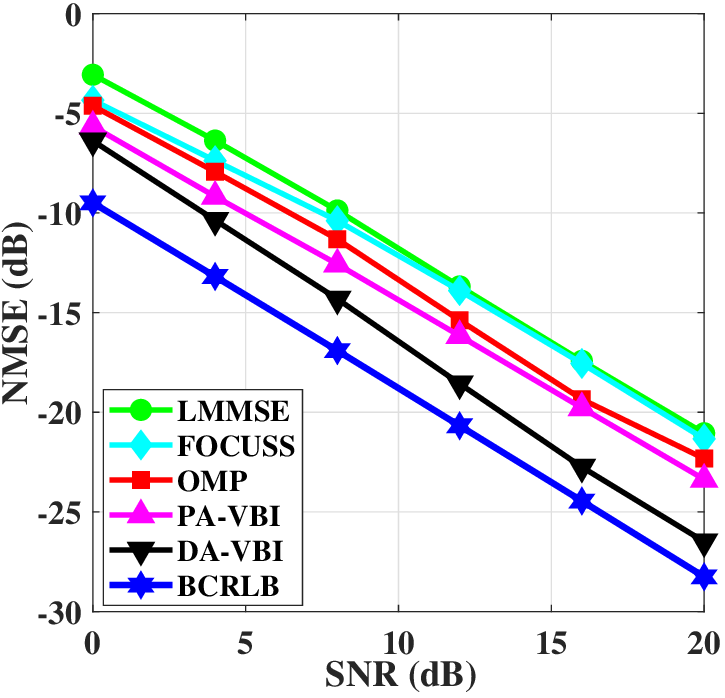}} 
\subfloat[]{\label{6}\includegraphics[width=60mm,height=50mm]{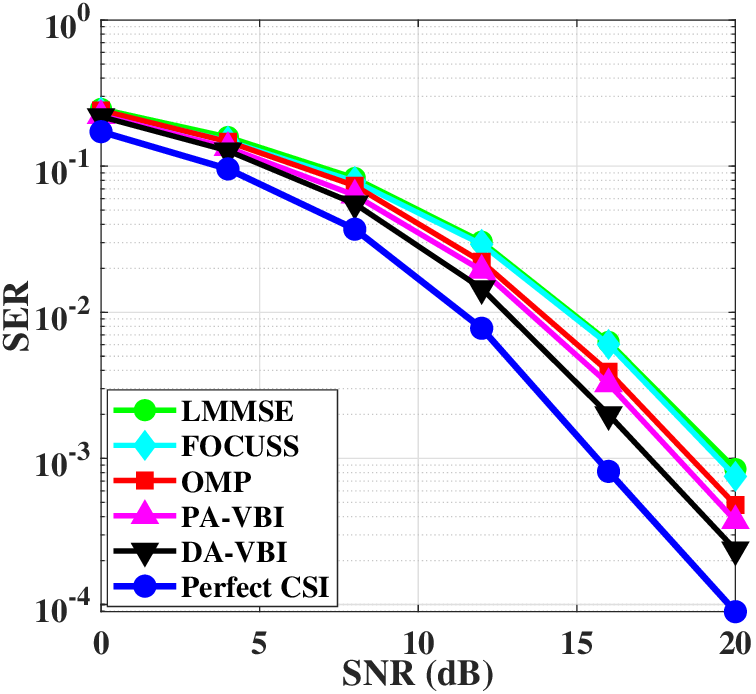}}
\caption{AP-STS MIMO DCO-OTFS VLC system, demonstrates (a) SER versus SNR performance with higher-order modulation. (b) NMSE versus SNR performance;  (c) SER versus SNR performance.}
\label{c}
\vspace{-2mm}
\end{figure*}

\begin{figure*}[ht]
\centering
\captionsetup[subfigure]{justification=centering}
\subfloat[]{\label{7}\includegraphics[width=60mm,height=50mm]{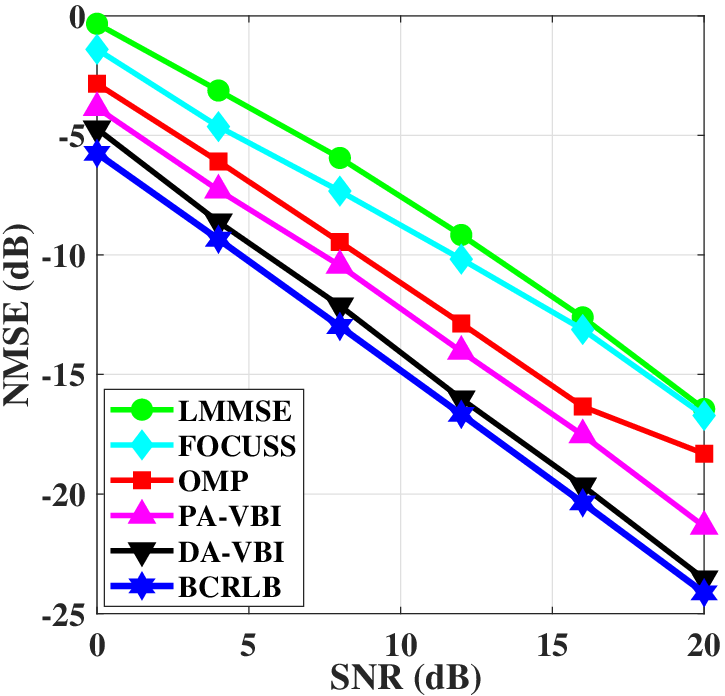}} \hspace{10mm}
\subfloat[]{\label{8}\includegraphics[width=60mm,height=50mm]{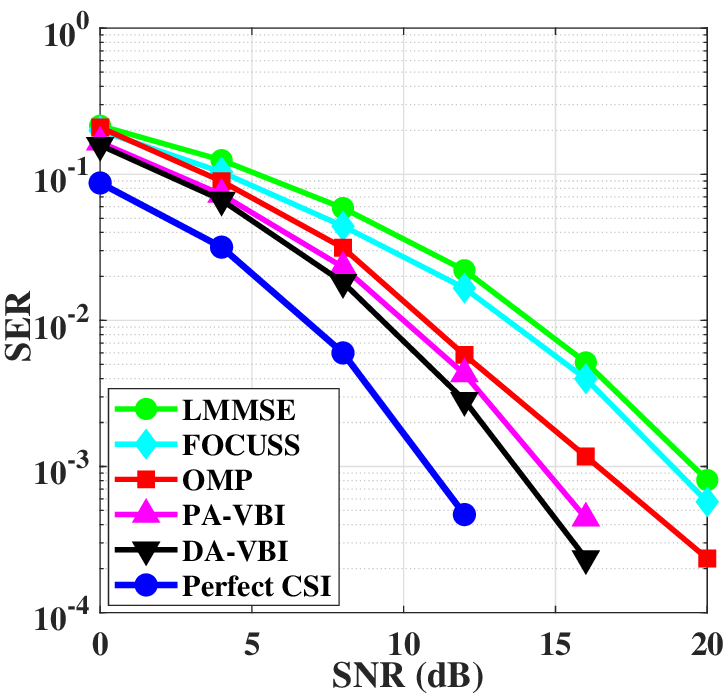}}
\caption{AP-STS MIMO DCO-OTFS VLC system with fractional Doppler, demonstrates (a) NMSE versus SNR performance; (b) SER versus SNR performance.}
\label{d}
\vspace{-2mm}
\end{figure*}

Fig. \ref{a}\subref{1} illustrates the NMSE performance of the proposed PA-VBI and DA-VBI estimators in comparison with the benchmark methods. The NMSE is defined as $\mathrm{NMSE}=\frac{||\widehat{\mathbf{h}}-\mathbf{h}||_2^2}{||\mathbf{h}||_2^2}.$
As shown in Fig. \ref{a}\subref{1}, the VBI-based PA-VBI and DA-VBI schemes provide lower NMSE than OMP, FOCUSS, and LMMSE for the AP-STS MIMO DCO-OTFS VLC system. The relatively inferior performance of OMP is mainly due to its dependence on an empirically chosen stopping rule. In comparison, FOCUSS is affected by convergence issues and high sensitivity to the regularization parameter, which limits its robustness \cite{saxena2023sparse}. Since the conventional LMMSE estimator does not exploit DD-domain CSI sparsity, it yields the weakest NMSE performance among the considered methods. Overall, the non-Bayesian sparse recovery schemes, namely OMP and FOCUSS, are less dependable than the VBI-based estimators for the above reasons. Among all the schemes, DA-VBI offers the best performance. This improvement is attributed to its use of the data estimates obtained from \eqref{fg1}, while still operating with limited pilot overhead. It is also observed that DA-VBI approaches the BCRLB in the high-SNR regime, despite not requiring prior knowledge of the channel covariance matrix, which is generally needed in conventional LMMSE processing. In addition, it does not assume prior knowledge of the sparse support. These features demonstrate the practical relevance of DA-VBI for DCO-OTFS implementations, where reliable prior information is often unavailable. Its joint CSI estimation and detection capability further improves performance by utilizing the large number of data symbols together with the relatively small number of pilots, resulting in behavior that closely follows the BCRLB at high SNR.

Fig. \ref{a}\subref{2} further shows the SER performance obtained when detecting the superimposed data symbols using the CSI delivered by each estimator. In particular, the SER is evaluated for the data symbols superimposed on the pilots by using the CSI obtained from the considered estimation schemes, and the results are also compared with those of an ideal receiver equipped with perfect CSI. As expected, the VBI-based methods, namely PA-VBI and DA-VBI, achieve lower SER than the non-VBI baselines OMP and FOCUSS, which is consistent with the NMSE behavior shown in Fig. \ref{a}\subref{1}. Moreover, DA-VBI again provides the best SER performance, with its curve lying close to that of the perfect-CSI benchmark. These results verify the effectiveness of the proposed DA-VBI scheme in generating highly accurate CSI estimates for reliable data detection.

Fig. \ref{a}\subref{3} shows the NMSE versus the pilot length $K_2$ for the proposed sparse CSI estimation schemes in the AP-STS MIMO DCO-OTFS VLC system at $\mathrm{SNR}=10$ dB and $20$ dB. As expected, the NMSE decreases monotonically with increasing $K_2$, which reflects the benefit of additional training in improving estimation accuracy. In particular, a larger $K_2$ provides more observations and hence reduces the estimation error \cite{kay1993fundamentals}. Among the considered methods, DA-VBI consistently achieves the lowest NMSE, which highlights the advantage of jointly exploiting the data symbols. In the proposed framework, $MK_2N_t$ pilot symbols are embedded in a block containing $MN_aN_t$ symbols, resulting in a pilot overhead of $\rho=\frac{K_2}{N_a}$. In contrast, the embedded-pilot (EP) scheme incurs a considerably larger overhead, approximately given by $\rho_{\mathrm{EP}}\approx \frac{(N_t M_\tau + M_\tau + N_t)(2N_{\nu}+1)}{MN_aN_t}$ \cite{raviteja2019embedded}. For the adopted parameters with $K_2=4$, the corresponding values are $\rho_{\mathrm{EP}}=0.42$ and $\rho=0.13$, confirming the superior bandwidth efficiency of the proposed design.

Fig. \ref{c}\subref{4} compares the SER of DA-VBI, PA-VBI, and OMP for $4$-, $16$-, and $64$-QAM signaling. In each case, DA-VBI yields the lowest SER, indicating that it supports more reliable detection under the CSI it produces. These results further demonstrate that the proposed DA-VBI remains effective over different modulation orders and consistently outperforms both PA-VBI and OMP across the considered SNR range.

To further demonstrate the effectiveness of the proposed framework, the DA-VBI and PA-VBI schemes are also evaluated under an additional simulation setting with $\Delta f=480$ kHz, $M=32$, $N=64$, $K_2=6$, $L=8$, $L_p=4$, $M_\tau=8$, and $N_\nu=7$, while all remaining parameters are kept unchanged. Fig. \ref{c}\subref{5} compares the NMSE of the proposed estimators with representative benchmark methods for the AP-STS MIMO DCO-OTFS VLC system, where DA-VBI again consistently provides the best performance. The corresponding SER results are reported in Fig. \ref{c}\subref{6}, which shows that DA-VBI attains significantly lower SER than PA-VBI, OMP, and FOCUSS. This gain arises from the improved CSI quality delivered by DA-VBI, which enables performance close to that of a receiver equipped with perfect CSI.

Figs. \ref{d}\subref{7} and \ref{d}\subref{8} present the NMSE and SER of the considered algorithms for AP-STS MIMO DCO-OTFS VLC links with $G_\nu=32$, thereby accounting for fractional Doppler effects. The results indicate that DA-VBI consistently outperforms the competing estimators, owing to its superior CSI quality and the resulting improvement in detection reliability. In addition, a larger number of Doppler bins $G_\nu$ refines the DD-grid resolution, which further enhances both estimation and detection performance.
\vspace{-3mm}
\section{Conclusions}
This paper developed an orthogonal AP-STS framework for CP-assisted MIMO DCO-OTFS VLC links operating over doubly selective channels. By jointly superimposing and affine-precoding the pilot and data matrices in the DD domain, a unified end-to-end DD-domain input-output relation was established. In addition, the use of orthogonal precoders at each PD enabled effective pilot-data separation while suppressing mutual interference. Based on this model, an EM-based PA-VBI algorithm was formulated for CSI estimation, followed by a DA-VBI scheme that iteratively refines the CSI and performs data detection by reusing the detected symbols as virtual pilots. BCRLB expressions were also derived for the considered system. Numerical results confirm that the proposed framework achieves lower NMSE, reduced pilot overhead, and improved SER relative to the benchmark schemes.
\vspace{-5mm}
\section*{Appendix A}
We examine the variational lower bound ${\mathcal {L}}(q(\pmb \Psi))$ as defined in \eqref{eq:elbo_nsbl_otfs}. By incorporating the expression from \eqref{eq:mf_factorization_nsbl_otfs} into this formulation, we arrive at
\vspace{-2mm}
\begin{align} 
& {\mathcal {L}}(q(\pmb \Psi)) = \int \prod_{t} q\left(\pmb\Psi_t\right)\left \{\ln f\left ({\mathbf{y}}^{\mathrm{dd},p}, {\pmb\Psi}\right)-\sum_{t} \ln q\left (\pmb\Psi_t\right)\right \} d \pmb\Psi \nonumber \\ & 
= \int q\left (\pmb\Psi_t\right)\left \{{{\int \ln f\left ({\mathbf{y}}^{\mathrm{dd},p}, {\pmb\Psi}\right) \prod _{j \neq t} q\left (\pmb{\Psi}_j\right) d {\pmb\Psi}_{\sim t}}}\right \} d \pmb\Psi_t \nonumber \\ & 
-\int q\left (\pmb\Psi_t\right) \ln q\left (\pmb\Psi_t\right) d \pmb\Psi_t+\text { constant term}~ \nonumber \\
&= \int q\left (\pmb\Psi_t\right) \ln \tilde{f}\left ({\mathbf{y}}^{\mathrm{dd},p}, {\pmb\Psi}_t\right) d \pmb\Psi_t-\int q\left (\pmb\Psi_t\right) \ln q\left (\pmb\Psi_t\right) d \pmb\Psi_t \nonumber \\ & 
+\text { constant term}~ \nonumber \\ & 
= -\text{KL}\left (q\left (\pmb\Psi_i\right)|| \tilde{f}\left ({\mathbf{y}}^{\mathrm{dd},p}, {\pmb\Psi}_t\right)\right)+\text{constant term},
\end{align}
where $\ln \tilde{f}\left ({\mathbf{y}}^{\mathrm{dd},p}, {\pmb\Psi}_t\right)$ is defined as
\vspace{-2mm}
\begin{equation} 
\ln \tilde{f}\left ({\mathbf{y}}^{\mathrm{dd},p}, {\pmb\Psi}_t\right)=\int \ln f\left ({\mathbf{y}}^{\mathrm{dd},p}, {\pmb\Psi}\right) \prod _{j \neq t} q\left ({\pmb\Psi}_j\right) d {{\pmb\Psi}}_{\sim t}. 
\end{equation}
Here, the integration is over all $q\left ({\pmb\Psi}_j\right), j \neq t$. The constant term is independent of $\pmb{\Psi}_t$. Maximizing ${\mathcal {L}}(q(\pmb \Psi))$ is hence equivalent to minimizing $\text{KL}\left (q\left (\pmb\Psi_t\right)|| \tilde{f}\left ({\mathbf{y}}^{\mathrm{dd},p}, {\pmb\Psi}_t\right)\right)$, which reaches its minimum when $q\left ({\pmb\Psi}_t\right) = \tilde{f}\left ({\mathbf{y}}^{\mathrm{dd},p}, {\pmb\Psi}_t\right)$, yielding
\vspace{-2mm}
\begin{align} 
q\left (\pmb\Psi_t\right)& = \tilde{f}\left ({\mathbf{y}}^{\mathrm{dd},p}, {\pmb\Psi}_t\right) \nonumber \\ & 
=\frac {1}{c} \exp \left \{{{\int \ln f\left ({\mathbf{y}}^{\mathrm{dd},p}, {\pmb\Psi}\right) \prod _{j \neq t} q\left ({\pmb\Psi}_j\right) d {\pmb\Psi}_{\sim t}}}\right \} \nonumber \\ & 
\propto \exp \Bigg \{\mathbb {E}_{\sim q\left ({\pmb\Psi}_t\right)} \bigg (\ln f\left ({\mathbf{y}}^{\mathrm{dd},p}, {\pmb\Psi}\right)\bigg)\Bigg \}, 
\end{align}
where the expectation is computed with respect to all $q\left ({\pmb\Psi}_j\right), j \neq t$, and $c$ is the normalization constant.
\vspace{-4mm}

\section*{Appendix B}
This appendix derives $\big(\mathbb{E}_{q^{(i)}(\mathbf{h})}(\widetilde{\mathbf{H}})\big)^{H}$ and $\mathbb{E}_{q^{(i)}(\mathbf{h})}(\widetilde{\mathbf{H}}^{H}\widetilde{\mathbf{H}})$. From the refined DD-domain channel model, define the deterministic matrix associated with the $(i,j)$th delay-Doppler tap as $\mathbf{T}^{(i,j)} \triangleq \mathbf{P}_{\mathrm{rx}}(\bar{\mathbf{\Pi}})^{i}\bar{\mathbf{\Delta}}_{i,j}\mathbf{P}_{\mathrm{tx}}.$
Let $\widetilde{\mathbf{T}}^{(i,j)}_{r,t}$ denote the deterministic matrix formed by placing $\mathbf{T}^{(i,j)}$ in the $(r,t)$th block position of $\widetilde{\mathbf{H}}^{\mathrm{dd}}$ and applying the same Kronecker lifting induced by $\widetilde{\mathbf{H}}=\mathbf{I}_{K_1}\otimes\widetilde{\mathbf{H}}^{\mathrm{dd}}$. Then,
\vspace{-2mm}
\begin{equation}
\widetilde{\mathbf{H}}=\sum_{r=1}^{N_r}\sum_{t=1}^{N_t}\sum_{i=0}^{M_\tau-1}\sum_{j=0}^{G_\nu-1}
h_{i,j,r,t}\widetilde{\mathbf{T}}^{(i,j)}_{r,t}.
\label{eq:app_h_tilde_expansion}
\end{equation}
Using \eqref{eq:app_h_tilde_expansion}, we obtain
\vspace{-2mm}
\begin{align}
\big(\mathbb{E}_{q^{(i)}(\mathbf{h})}(\widetilde{\mathbf{H}})\big)^{H}
&=\mathlarger{\sum_{r,t,i,j}}
\mathbb{E}_{q^{(i)}(\mathbf{h})}(h_{i,j,r,t})^{*}
\big(\widetilde{\mathbf{T}}^{(i,j)}_{r,t}\big)^{H},
\label{eq:app_E_h_tilde_H}\\
\mathbb{E}_{q^{(i)}(\mathbf{h})}\big(\widetilde{\mathbf{H}}^{H}\widetilde{\mathbf{H}}\big) &=\mathlarger{\sum_{r,t,i,j}}~\mathlarger{\sum_{r',t',i',j'}}
\mathbb{E}_{q^{(i)}(\mathbf{h})}\big(h_{i,j,r,t}^{*}h_{i',j',r',t'}\big) \nonumber \\
&\times \big(\widetilde{\mathbf{T}}^{(i,j)}_{r,t}\big)^{H}\widetilde{\mathbf{T}}^{(i',j')}_{r',t'}.
\label{eq:app_E_hHt_h}
\end{align}
Let $\ell$ and $\ell'$ denote the composite indices in the stacked vector $\mathbf{h}$ corresponding to $(i,j,r,t)$ and $(i',j',r',t')$, respectively. Then the second-order moment is given by
\vspace{-2mm}
\begin{equation}
\mathbb{E}\big(h_{\ell}^{*}h_{\ell'}\big)=
\big[\pmb{\Sigma}_{\mathbf{h}}^{(i)}+\pmb{\mu}_{\mathbf{h}}^{(i)}\big(\pmb{\mu}_{\mathbf{h}}^{(i)}\big)^{H}\big]_{(\ell',\ell)},
\label{eq:app_second_moment_rule}
\end{equation}
where $\pmb{\mu}_{\mathbf{h}}^{(i)}=\mathbb{E}_{q^{(i)}(\mathbf{h})}(\mathbf{h})$ and
$\pmb{\Sigma}_{\mathbf{h}}^{(i)}=\mathbb{E}_{q^{(i)}(\mathbf{h})}\big((\mathbf{h}-\pmb{\mu}_{\mathbf{h}}^{(i)})(\mathbf{h}-\pmb{\mu}_{\mathbf{h}}^{(i)})^{H}\big)$.

\vspace{-4mm}

\section*{Appendix C} 
This appendix derives $\mathbb{E}_{q^{(i)}(\mathbf{s}^{d})}(\boldsymbol{\Phi})$, $\mathbb{E}_{q^{(i)}(\mathbf{s}^{d})}(\boldsymbol{\Phi}^{H}\boldsymbol{\Phi})$, and
$\mathrm{tr}\Big\{\mathbb{E}_{q^{(i)}(\mathbf{s}^{d})}\big(\boldsymbol{\Phi}\boldsymbol{\Sigma}_{\mathbf{h}}^{(i)}\boldsymbol{\Phi}^{H}\big)\Big\}$.
From \eqref{po1}, the sensing matrix has the block structure
\begin{equation}
\boldsymbol{\Phi}
=\begin{bmatrix}
\boldsymbol{\Omega}^{d}\\
\boldsymbol{\Omega}^{p}
\end{bmatrix},
\end{equation}
where $\boldsymbol{\Omega}^{p}$ is deterministic, while $\boldsymbol{\Omega}^{d}$ depends on the DD-domain data vector $\mathbf{s}^{d}$ through
\vspace{-2mm}
\begin{equation}
\boldsymbol{\omega}^{d}_{i,j,t}
=\big(\mathbf{I}_{K_{1}}\otimes \mathbf{P}_{\mathrm{rx}}(\bar{\boldsymbol{\Pi}})^{i}\bar{\boldsymbol{\Delta}}_{i,j}\mathbf{P}_{\mathrm{tx}}\big)\mathbf{s}^{d}_{t}
\triangleq \mathbf{B}_{i,j}\mathbf{s}^{d}_{t}.
\label{B.2}
\end{equation}
Therefore, the expectation of $\boldsymbol{\Phi}$ with respect to $q^{(i)}(\mathbf{s}^{d})$ is
\vspace{-2mm}
\begin{align}
\mathbb{E}_{q^{(i)}(\mathbf{s}^{d})}\big(\boldsymbol{\Phi}\big)
&=\begin{bmatrix}
\mathbb{E}_{q^{(i)}(\mathbf{s}^{d})}\big(\boldsymbol{\Omega}^{d}\big)\\[0.5mm]
\boldsymbol{\Omega}^{p}
\end{bmatrix}, \text{ and } \nonumber\\
\mathbb{E}_{q^{(i)}(\mathbf{s}^{d})}\big(\boldsymbol{\omega}^{d}_{i,j,t}\big)
&=\mathbf{B}_{i,j}\mathbb{E}_{q^{(i)}(\mathbf{s}^{d})}\big(\mathbf{s}^{d}_{t}\big).
\label{B.3}
\end{align}
Since $\boldsymbol{\Phi}$ is obtained by vertical concatenation, it follows that
$\boldsymbol{\Phi}^{H}\boldsymbol{\Phi}
=(\boldsymbol{\Omega}^{d})^{H}\boldsymbol{\Omega}^{d}
+(\boldsymbol{\Omega}^{p})^{H}\boldsymbol{\Omega}^{p},$ and hence
\vspace{-2mm}
\begin{equation}
\mathbb{E}_{q^{(i)}(\mathbf{s}^{d})}\big(\boldsymbol{\Phi}^{H}\boldsymbol{\Phi}\big)
=\mathbb{E}_{q^{(i)}(\mathbf{s}^{d})}\big((\boldsymbol{\Omega}^{d})^{H}\boldsymbol{\Omega}^{d}\big)+
(\boldsymbol{\Omega}^{p})^{H}\boldsymbol{\Omega}^{p}.
\label{B.4}
\end{equation}
To compute $\mathbb{E}_{q^{(i)}(\mathbf{s}^{d})}\big((\boldsymbol{\Omega}^{d})^{H}\boldsymbol{\Omega}^{d}\big)$, we use the second-order moment identity
\vspace{-2mm}
\begin{equation}
\mathbb{E}_{q^{(i)}(\mathbf{s}^{d})}\big(\mathbf{s}^{d}(\mathbf{s}^{d})^{H}\big)
=\boldsymbol{\Sigma}_{\mathbf{s}}^{(i)}+
\mathbf{u}_{\mathbf{s}}^{(i)}\big(\mathbf{u}_{\mathbf{s}}^{(i)}\big)^{H},
\label{B.6}
\end{equation}
where $\mathbf{u}_{\mathbf{s}}^{(i)}=\mathbb{E}_{q^{(i)}(\mathbf{s}^{d})}(\mathbf{s}^{d})$ and
$\boldsymbol{\Sigma}_{\mathbf{s}}^{(i)}=\mathbb{E}_{q^{(i)}(\mathbf{s}^{d})}\big((\mathbf{s}^{d}-\mathbf{u}_{\mathbf{s}}^{(i)})(\mathbf{s}^{d}-\mathbf{u}_{\mathbf{s}}^{(i)})^{H}\big)$.
Equivalently, for $1\le n,m\le Q$,
\vspace{-2mm}
\begin{equation}
\mathbb{E}_{q^{(i)}(\mathbf{s}^{d})}\big((s^{d}_{n})^{*}s^{d}_{m}\big)
=
\left[
\boldsymbol{\Sigma}_{\mathbf{s}}^{(i)}+
\mathbf{u}_{\mathbf{s}}^{(i)}\big(\mathbf{u}_{\mathbf{s}}^{(i)}\big)^{H}
\right]_{(m,n)}.
\label{B.7}
\end{equation}
In many implementations, one may further approximate $\boldsymbol{\Sigma}_{\mathbf{s}}^{(i)}\approx \mathbf{0}$ to reduce computational cost. Under this approximation, \eqref{B.7} reduces to
\vspace{-2mm}
\begin{equation}
\mathbb{E}_{q^{(i)}(\mathbf{s}^{d})}\big((s^{d}_{n})^{*}s^{d}_{m}\big)
\approx
\big[\mathbf{u}_{\mathbf{s}}^{(i)}\big]_{(n)}^{*}\big[\mathbf{u}_{\mathbf{s}}^{(i)}\big]_{(m)}.
\end{equation}
Finally, using the cyclic property of the trace operator, the term
$\mathrm{tr}\Big\{\mathbb{E}_{q^{(i)}(\mathbf{s}^{d})}\big(\boldsymbol{\Phi}\boldsymbol{\Sigma}_{\mathbf{h}}^{(i)}\boldsymbol{\Phi}^{H}\big)\Big\}$
can be rewritten as
\vspace{-2mm}
\begin{align}
&\mathrm{tr}\Big\{\mathbb{E}_{q^{(i)}(\mathbf{s}^{d})}\big(\boldsymbol{\Phi}\boldsymbol{\Sigma}_{\mathbf{h}}^{(i)}\boldsymbol{\Phi}^{H}\big)\Big\}
=
\mathrm{tr}\Big\{\mathbb{E}_{q^{(i)}(\mathbf{s}^{d})}\big(\boldsymbol{\Phi}^{H}\boldsymbol{\Phi}\boldsymbol{\Sigma}_{\mathbf{h}}^{(i)}\big)\Big\} \nonumber\\
&=
\mathrm{tr}\Big\{\mathbb{E}_{q^{(i)}(\mathbf{s}^{d})}\big(\boldsymbol{\Phi}^{H}\boldsymbol{\Phi}\big)\boldsymbol{\Sigma}_{\mathbf{h}}^{(i)}\Big\},
\label{B.8}
\end{align}
where $\mathbb{E}_{q^{(i)}(\mathbf{s}^{d})}\big(\boldsymbol{\Phi}^{H}\boldsymbol{\Phi}\big)$ is obtained from \eqref{B.4}-\eqref{B.7}.

\bibliographystyle{IEEEtran}
\bibliography{citation}

@article{zhong2020orthogonal,
  title={Orthogonal time-frequency multiplexing with {2D} {Hermitian} symmetry for optical-wireless communications},
  author={Zhong, Jie and Zhou, Ji and Liu, Weiping and Qin, Jiayin},
  journal={IEEE Photonics Journal},
  volume={12},
  number={2},
  pages={1--10},
  year={2020},
  publisher={IEEE}
}

@article{zheng2021dco,
  title={{DCO-OTFS}-based full-duplex relay-assisted visible light communications},
  author={Zheng, Di and Zhang, Hongming and Song, Jian},
  journal={Optics Express},
  volume={29},
  number={25},
  pages={41323--41332},
  year={2021},
  publisher={Optical Society of America}
}

@article{sharma2023hyperparameter,
  title={Hyperparameter-free {RFF}-based post-distorter for {OTFS VLC} system},
  author={Sharma, Anupma and Mitra, Rangeet and Krejcar, Ondrej and Choi, Kwonhue and Dobrovolny, Michal and Bhatia, Vimal},
  journal={IEEE Photonics Journal},
  volume={15},
  number={2},
  pages={1--7},
  year={2023},
  publisher={IEEE}
}

@article{xu2023optical,
  title={Optical {OTFS} is capable of improving the bandwidth-, power-and energy-efficiency of optical {OFDM}},
  author={Xu, Chao and Zhang, Xiaoyu and Petropoulos, Periklis and Sugiura, Shinya and Maunder, Robert G and Yang, Lie-Liang and Wang, Zhaocheng and Yuan, Jinghong and Haas, Harald and Hanzo, Lajos},
  journal={IEEE Transactions on Communications},
  volume={72},
  number={2},
  pages={938--953},
  year={2023},
  publisher={IEEE}
}

@article{cai2025power,
  title={A Power-and Spectrum-Efficient Underwater Wireless Optical Communication System based on a Hierarchical Pre-Distorted {LACO-OTFS} Scheme},
  author={Cai, Chengye and Hu, Xiaoqi and Du, Zihao and Zhu, Boyu and Liao, Wendong and Ma, Xiaoxu and Chen, Baokang and Chen, Qingrui and Ge, Wenmin and Song, Guangbin and others},
  journal={Journal of Lightwave Technology},
  year={2025},
  publisher={IEEE}
}

@article{chen2025optical,
  title={Optical {OTFS} Modulation for Free Space Optical-Based {LEO} Satellite Communication Systems},
  author={Chen, Wenbin and Ju, Cheng and Yuan, Tianxing and Li, Jin and Zhang, Min and Wang, Danshi},
  journal={IEEE Photonics Technology Letters},
  year={2025},
  publisher={IEEE}
}

@inproceedings{liao2023sparse,
  title={Sparse {Bayesian} Learning-based Channel Estimation for Indoor {OTFS} Visible Light Communication},
  author={Liao, Yuxuan and Pei, Jianhua and Dai, Weijie and Song, Jian and Dong, Yuhan},
  booktitle={2023 Asia Communications and Photonics Conference/2023 International Photonics and Optoelectronics Meetings (ACP/POEM)},
  pages={1--5},
  year={2023},
  organization={IEEE}
}

@article{saxena2023sparse,
  title={Sparse channel estimation for visible light optical {OFDM} systems relying on {Bayesian} learning},
  author={Saxena, Shubham and Srivastava, Suraj and Sharma, Saurabh and Jagannatham, Aditya K and Hanzo, Lajos},
  journal={IEEE Open Journal of the Communications Society},
  volume={4},
  pages={2062--2079},
  year={2023},
  publisher={IEEE}
}

@inproceedings{hadani2017orthogonal,
  title={Orthogonal time frequency space modulation},
  author={Hadani, Ronny and Rakib, Shlomo and Tsatsanis, Michail and Monk, Anton and Goldsmith, Andrea J and Molisch, Andreas F and Calderbank, R},
  booktitle={2017 IEEE wireless communications and networking conference (WCNC)},
  pages={1--6},
  year={2017},
  organization={IEEE}
}

@article{raviteja2019otfs,
  title={{OTFS} performance on static multipath channels},
  author={Raviteja, Patchava and Viterbo, Emanuele and Hong, Yi},
  journal={IEEE Wireless Communications Letters},
  volume={8},
  number={3},
  pages={745--748},
  year={2019},
  publisher={IEEE}
}

@article{wipf2007empirical,
  title={An empirical {Bayesian} strategy for solving the simultaneous sparse approximation problem},
  author={Wipf, David P and Rao, Bhaskar D},
  journal={IEEE Transactions on Signal Processing},
  volume={55},
  number={7},
  pages={3704--3716},
  year={2007},
  publisher={IEEE}
}

@inproceedings{sinha2021otfs,
  title={{OTFS} modulation in {Dual-LED} indoor visible light communication systems},
  author={Sinha, Sujata and Chockalingam, A},
  booktitle={2021 IEEE 94th Vehicular Technology Conference (VTC2021-Fall)},
  pages={1--7},
  year={2021},
  organization={IEEE}
}

@article{estrada2019superimposed,
  title={Superimposed training-based channel estimation for {MISO} {optical-OFDM VLC}},
  author={Estrada-Jim{\'e}nez, Juan Carlos and Guzm{\'a}n, Borja Genov{\'e}s and Garc{\'\i}a, M Julia Fern{\'a}ndez-Getino and Jim{\'e}nez, V{\'\i}ctor P Gil},
  journal={IEEE Transactions on Vehicular Technology},
  volume={68},
  number={6},
  pages={6161--6166},
  year={2019},
  publisher={IEEE}
}

@article{muntane2024optimal,
  title={Optimal estimation of frequency-selective channels in {OFDM}-based superimposed training schemes},
  author={Muntane, Ignasi Pique and Garc{\'\i}a, M Julia Fern{\'a}ndez-Getino},
  journal={IEEE Transactions on Vehicular Technology},
  volume={73},
  number={11},
  pages={15969--15982},
  year={2024},
  publisher={IEEE}
}

@inproceedings{mishra2021iterative,
  title={Iterative channel estimation and data detection in {OTFS} using superimposed pilots},
  author={Mishra, Himanshu B and Singh, Prem and Prasad, Abhishek K and Budhiraja, Rohit},
  booktitle={2021 IEEE International Conference on Communications Workshops (ICC Workshops)},
  pages={1--6},
  year={2021},
  organization={IEEE}
}

@article{yuan2021data,
  title={Data-aided channel estimation for {OTFS} systems with a superimposed pilot and data transmission scheme},
  author={Yuan, Weijie and Li, Shuangyang and Wei, Zhiqiang and Yuan, Jinhong and Ng, Derrick Wing Kwan},
  journal={IEEE Wireless Communications Letters},
  volume={10},
  number={9},
  pages={1954--1958},
  year={2021},
  publisher={IEEE}
}

@inproceedings{ramachandran2018mimo,
  title={{MIMO-OTFS} in high-{Doppler} fading channels: Signal detection and channel estimation},
  author={Ramachandran, M Kollengode and Chockalingam, A},
  booktitle={2018 IEEE Global Communications Conference (GLOBECOM)},
  pages={206--212},
  year={2018},
  organization={IEEE}
}

@article{wang2025spectrally,
  title={Spectrally Efficient Optical {OTFS} with Enhanced Joint Delay-{Doppler} Index Modulation},
  author={Wang, Zhen and Wang, Huiqin and Tang, Qihan and Peng, Qingbin and Zhang, Yue and Cao, Minghua},
  journal={IEEE Photonics Technology Letters},
  year={2025},
  publisher={IEEE}
}

@article{tran2008orthogonal,
  title={Orthogonal affine precoding and decoding for channel estimation and source detection in {MIMO} frequency-selective fading channels},
  author={Tran, Nguyen Nam and Pham, Duong H and Tuan, Hoang Duong and Nguyen, Ha H},
  journal={IEEE Transactions on Signal Processing},
  volume={57},
  number={3},
  pages={1151--1162},
  year={2008},
  publisher={IEEE}
}

@inproceedings{sinha2021quad,
  title={Quad-{LED} {OTFS} modulation in indoor visible light communication systems},
  author={Sinha, Sujata and Chockalingam, A},
  booktitle={2021 IEEE Global Communications Conference (GLOBECOM)},
  pages={1--6},
  year={2021},
  organization={IEEE}
}

@book{kay1993fundamentals,
  title={Fundamentals of statistical signal processing: estimation theory},
  author={Kay, Steven M},
  year={1993},
  publisher={Prentice-Hall, Inc.}
}

@article{h1,
  author={Zhang, Xiaoyu and Babar, Zunaira and Petropoulos, Periklis and Haas, Harald and Hanzo, Lajos},
  journal={IEEE Communications Surveys \& Tutorials}, 
  title={The Evolution of Optical {OFDM}}, 
  year={2021},
  volume={23},
  number={3},
  pages={1430-1457},
  doi={10.1109/COMST.2021.3065907}}

@article{8123892,
  author={Wang, Jingjing and Jiang, Chunxiao and Zhang, Haijun and Zhang, Xin and Leung, Victor C. M. and Hanzo, Lajos},
  journal={IEEE Transactions on Vehicular Technology}, 
  title={Learning-Aided Network Association for Hybrid Indoor {LiFi-WiFi} Systems}, 
  year={2018},
  volume={67},
  number={4},
  pages={3561-3574},
  doi={10.1109/TVT.2017.2778345}}

@article{chen2016adaptive,
  title={Adaptive statistical {Bayesian} {MMSE} channel estimation for visible light communication},
  author={Chen, Xianyu and Jiang, Ming},
  journal={IEEE Transactions on Signal Processing},
  volume={65},
  number={5},
  pages={1287--1299},
  year={2016},
  publisher={IEEE}
}

@article{schulze2016frequency,
  title={Frequency-domain simulation of the indoor wireless optical communication channel},
  author={Schulze, Henrik},
  journal={IEEE Transactions on Communications},
  volume={64},
  number={6},
  pages={2551--2562},
  year={2016},
  publisher={IEEE}
}

@article{zhou2014impact,
  title={Impact analyses of high-order light reflections on indoor optical wireless channel model and calibration},
  author={Zhou, Zhou and Chen, Chunyi and Kavehrad, Mohsen},
  journal={Journal of Lightwave Technology},
  volume={32},
  number={10},
  pages={2003--2011},
  year={2014},
  publisher={OSA}
}

@article{van2021deep,
  title={Deep learning-aided optical {IM/DD} {OFDM} approaches the throughput of {RF-OFDM}},
  author={Van Luong, Thien and Zhang, Xiaoyu and Xiang, Luping and Hoang, Tiep M and Xu, Chao and Petropoulos, Periklis and Hanzo, Lajos},
  journal={IEEE Journal on Selected Areas in Communications},
  volume={40},
  number={1},
  pages={212--226},
  year={2021},
  publisher={IEEE}
}

@article{g1,
  author={Hei, Yongqiang and Kou, Yanchun and Shi, Guangming and Li, Wentao and Gu, Huaxi},
  journal={IEEE Transactions on Vehicular Technology}, 
  title={Energy-Spectral Efficiency Tradeoff in {DCO-OFDM} Visible Light Communication System}, 
  year={2019},
  volume={68},
  number={10},
  pages={9872-9882},
  doi={10.1109/TVT.2019.2934132}}

@article{saxena2025multiple,
  author={Saxena, Shubham and Sharma, Saurabh and Srivastava, Suraj and Jagannatham, Aditya K. and Hanzo, Lajos},
  journal={IEEE Transactions on Vehicular Technology}, 
  title={Multiple Measurement Vector Based {Bayesian} Learning for Simultaneously Sparse Time/Delay-Domain Channel Estimation in {ADO-OFDM} Visible Light Systems}, 
  year={2025},
  volume={},
  number={},
  pages={1-16},
  doi={10.1109/TVT.2025.3639320}}

@article{srivastava2021bayesian,
  title={Bayesian learning aided sparse channel estimation for orthogonal time frequency space modulated systems},
  author={Srivastava, Suraj and Singh, Rahul Kumar and Jagannatham, Aditya K and Hanzo, Lajos},
  journal={IEEE Transactions on Vehicular Technology},
  volume={70},
  number={8},
  pages={8343--8348},
  year={2021},
  publisher={IEEE}
}

@article{raviteja2019embedded,
  title={Embedded pilot-aided channel estimation for {OTFS} in {delay-Doppler} channels},
  author={Raviteja, Patchava and Phan, Khoa T and Hong, Yi},
  journal={IEEE Transactions on Vehicular Technology},
  volume={68},
  number={5},
  pages={4906--4917},
  year={2019},
  publisher={IEEE}
}

@article{li2020new,
  title={A new path division multiple access for the massive {MIMO-OTFS} networks},
  author={Li, Muye and Zhang, Shun and Gao, Feifei and Fan, Pingzhi and Dobre, Octavia A},
  journal={IEEE Journal on Selected Areas in Communications},
  volume={39},
  number={4},
  pages={903--918},
  year={2020},
  publisher={IEEE}
}

@article{li2023doubly,
  title={Doubly-iterative sparsified {MMSE} turbo equalization for {OTFS} modulation},
  author={Li, Haotian and Yu, Qiyue},
  journal={IEEE Transactions on Communications},
  volume={71},
  number={3},
  pages={1336--1351},
  year={2023},
  publisher={IEEE}
}

@article{srivastava2022otfs,
  title={{OTFS} transceiver design and sparse doubly-selective {CSI} estimation in analog and hybrid beamforming aided mmWave {MIMO} systems},
  author={Srivastava, Suraj and Singh, Rahul Kumar and Jagannatham, Aditya K and Chockalingam, A and Hanzo, Lajos},
  journal={IEEE Transactions on Wireless Communications},
  volume={21},
  number={12},
  pages={10902--10917},
  year={2022},
  publisher={IEEE}
}

@article{liu2020uplink,
  title={Uplink-aided high mobility downlink channel estimation over massive {MIMO-OTFS} system},
  author={Liu, Yushan and Zhang, Shun and Gao, Feifei and Ma, Jianpeng and Wang, Xianbin},
  journal={IEEE Journal on Selected Areas in Communications},
  volume={38},
  number={9},
  pages={1994--2009},
  year={2020},
  publisher={IEEE}
}

@article{wang2022joint,
  title={Joint {Bayesian} channel estimation and data detection for {OTFS} systems in {LEO} satellite communications},
  author={Wang, Xueyang and Shen, Wenqian and Xing, Chengwen and An, Jianping and Hanzo, Lajos},
  journal={IEEE Transactions on Communications},
  volume={70},
  number={7},
  pages={4386--4399},
  year={2022},
  publisher={IEEE}
}
\end{document}